\documentclass{aa}

\newcommand{\cd}{$\mathrm{d}^{-1}$}

\usepackage{hyperref}
\usepackage{graphicx}
\usepackage{longtable}
\usepackage{multirow}
\usepackage{color}
\usepackage{soul}
\usepackage{placeins}
\usepackage{txfonts}
\usepackage[normalem]{ulem}

\begin{document}

   \title{Pulsational instability of pre-main-sequence models from accreting protostars}

   \subtitle{II. Modelling echelle diagrams of $\delta$ Scuti stars without rotational splitting}

   \author{T. Steindl\inst{1}, K. Zwintz\inst{1}, M. Müllner\inst{1}}
    
    \authorrunning{T. Steindl et al.}
    \titlerunning{Modelling Echelle diagrams of Delta Scuti Stars without rotational splitting}
    
   \institute{\inst{1}Institut f\"ur Astro- und Teilchenphysik, Universit\"at Innsbruck, Technikerstra{\ss}e 25, A-6020 Innsbruck, Austria\\
              \email{thomas.steindl@uibk.ac.at}  \\
              }

   \date{submitted 01.02.2022; accepted 25.05.2022}

 
  \abstract
   {The physics of early stellar evolution (e.g. accretion processes) is often not properly included in the calculations of pre-main-sequence models, leading to insufficient model grids and hence systematic errors in the results.}
   {We aim to investigate current and improved approaches for the asteroseismic modelling of pre-main-sequence $\delta$ Scuti stars. }
   {We calculated an extensive grid of pre-main-sequence models including the early accretion phase and used the resulting equilibrium models as input to calculate theoretical frequency spectra. These spectra were used to investigate different approaches in modelling echelle diagrams to find the most reliable methods. By applying Petersen diagrams, we present a simple algorithm to extract echelle diagrams from observed pulsation frequencies. }
   {We show that model grids with insufficient input physics and imperfect modelling approaches lead to underestimated uncertainties and systematic errors in the extracted stellar parameters. Our re-discussion of HD 139614 leads to different stellar parameters than the ones derived by \citet{murphy2021}. We performed a model comparison between this previous investigation and our results by applying the Akaike and Bayesian information criteria. While the results with regard to our ten-dimensional model are inconclusive, they show (very) strong evidence of a six-dimensional model with fixed accretion parameters (leading to almost identical stellar parameters to those of the ten-dimensional model) to be preferred over the model applied by \citet{murphy2021}. In general, our modelling approach can provide narrow constraints on the stellar parameters (i.e. $\Delta R \sim 0.05\,R_\odot$, $\Delta \log\,g \lesssim 0.01$, and $\Delta M_\star \sim 0.1\,M_\odot$). }
   {The extensively tested modelling approaches and automatic extraction of echelle diagrams should allow us to study many more pre-main-sequence $\delta$ Scuti stars in the future and lead to reliable stellar parameters. }

   \keywords{asteroseismology -- stars: oscillations (including pulsations) -- accretion, accretion disks -- stars: variables: delta Scuti -- stars: pre-main-sequence -- stars: protostars
               }

   \maketitle
%

\section{Introduction}
The evolution of intermediate-mass stars is often truncated to their main-sequence, red-giant, and asymptotic-giant-branch phases, and their final fate as white dwarfs. However, before stars are able to burn significant amounts of hydrogen in their cores, they undergo the phase of pre-main-sequence evolution. During this phase, which starts from the formation of the second Larson's core  (or second hydrastic core, see \citeauthor{Larson1969} \citeyear{Larson1969} and \citeauthor{Masunaga2000} \citeyear{Masunaga2000}), a multitude of physical processes shape the stars and, hence, their future fate. While these processes play a major role in stellar evolution (e.g. for the formation of magnetic fields, angular momentum transport, and planet formation), they are often ignored in model calculations. Many studies of early stellar evolution still rely on the pioneering works of \citet{Henyey1955}, \citet{Hayashi1961}, and \citet{Iben1965}, albeit ignoring first and foremost any accretion processes. While the outdated view of the pre-main-sequence evolution just assumes the initial models with huge radii, the addition of accretion processes in stellar modelling allows for a much more realistic view. This series of papers aims to combine such more realistic evolution models with asteroseismology, which is the study of pulsating stars. 

Pre-main-sequence stars with masses between $1.5$ and $3.5\,M_\odot$ are expected to become unstable to stellar pulsations \citep[e.g.][]{Steindl2021b} as $\delta$ Scuti \citep[e.g.][]{Zwintz2014} or slowly pulsating B-type stars \citep[e.g.][]{Gruber2012}. Pre-main-sequence $\delta$ Scuti stars are expected to have effective temperatures in the range from $6\,300$ to $10\,300$~K \citep{Steindl2021b} and pulsate in low-order pressure modes driven by the heat mechanism \citep{Dupret2005, Steindl2021b}. \citet{Zwintz2014} show that their pulsation characteristics are dependent on their relative evolutionary status, with stars further evolved towards the ZAMS pulsating at higher frequencies. \cite{bedding2020} present regularities in the pulsation spectra of mainly young main-sequence $\delta$ Scuti stars. Such regularities are often shown in echelle diagrams and allow the asteroseismic modelling of $\delta$ Scuti stars \citep[see e.g.][]{Bernabei2009, Chen2019, murphy2021, Kim2021}. 

In slowly rotating stars, echelle diagrams are expected to be much simpler compared to their fast-rotating counterparts, since pulsation modes are split according to the angular velocity of the stars \citep{Aerts2010}. 
It is possible to automatically infer an echelle diagram and, hence, identify the pulsation modes from the detected pulsation frequencies of $\delta$ Scuti stars. From that point on, it is a rather simple exercise to produce a grid of theoretical stellar evolution models and pulsation spectra and apply asteroseismic model approaches. 
Here, we present the limitations of this method if the underlying stellar grid insufficiently includes physics of early stellar evolution and if additional free parameters are ignored. In addition, the uncertainties might be underestimated and the stellar parameters might be extracted incorrectly.

In this work, we calculated an extensive grid of pre-main-sequence models in the mass range expected for $\delta$ Scuti stars, starting from the second hydrostatic core and including the accretion phase. In addition, we allowed for free parameters in the initial composition and mixing parameters needed to accurately describe possible evolutionary pathways of early stars. The computational set-up and calculation of the grid are described in Section \ref{sec:models}. In Section \ref{sec:modelling_approach}, we review, discuss, and introduce different approaches for the asteroseismic modelling of pre-main-sequence $\delta$ Scuti stars. These approaches were then extensively tested using our model grid, which we describe in Section \ref{sec:theoretical_echelle_diagrams}. We introduced a simple algorithm based on the results of the first paper of this series \citep{Steindl2021b} to automatically extract echelle diagrams, which we discuss in Section \ref{sec:automatic_echelle_diagrams}. We then performed asteroseismic modelling using the theoretical pulsation frequencies of our state-of-the-art pre-main-sequence grid on the star HD 139614, which was previously discussed by \citet{murphy2021}, and we offer a model comparison in Section \ref{sec:rediscussion}. We chose to re-discuss HD 139614 since it is very slowly rotating \citep{murphy2021} and, hence, a perfect specimen for the methodology described in this work. We conclude our results in Section \ref{sec:conlusion}.

\section{Computational set-up}
\label{sec:models}

\begin{table}
    \caption{Summary of the covered input physics.
    }
    \label{tab:summary_sample}
    \tabcolsep=0.16cm
    \begin{tabular*}{\linewidth}{llcc}
        \hline
        \noalign{\smallskip}
        param.   &  description & value range & step size \\
        \noalign{\smallskip}
        \hline
        \noalign{\smallskip}
        Z & metallicity & 0.006 - 0.024 & 0.002 \\
        Y & helium abundance & 0.204 - 0.234 & 0.01 \\
        $\alpha_{\rm MLT}$ & mixing length &  1.8 - 2.4 & 0.2 \\
        $M_\star$ & final mass & 1.5 - 2.5 & continuous \\
        $F$ & overshooting & 0-0.02 & continuous \\
        $D_{\rm mix}$  & envelope mixing & 0-100 ${\rm cm} {\rm s}^{-1}$ & continuous \\
        $\beta$  & fractional energy & 0.05-0.2 & continuous \\
        $M_{\rm outer}$  & fractional mass & 0.01-0.2 & continuous \\
        $\Dot{M}_0 $  & accretion rate & $10^{-7} - 10^{-5} \,M_\odot/{\rm yr}$  & continuous \\
        $\tau$  & e-folding time & 0.1 -1\,Myr  & continuous \\

    \noalign{\smallskip}
    \hline
    \end{tabular*}    
    \tablefoot{As each combination of the parameters Z, Y, and $\alpha_{\rm MLT}$ call for an initial model, we limited these parameters to periodic grids.}

 \end{table}
 
 \begin{figure*}
   \centering
   \includegraphics[width=\linewidth]{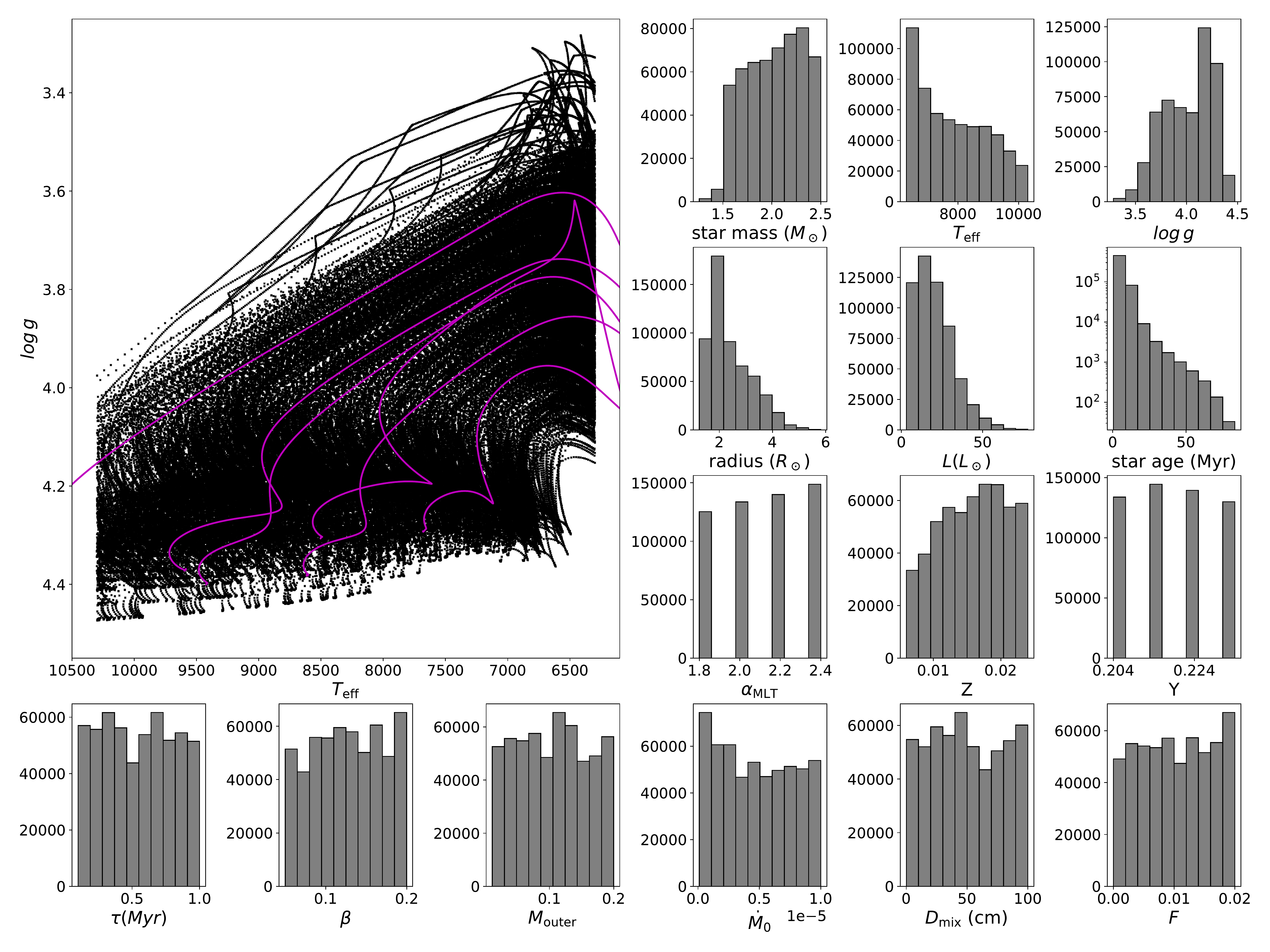}
      \caption{Parameters of 182\,547 distinct pre-main-sequence $\delta$ Scuti models. The top left panel shows a Kiel diagram with all models as black crosses and a few random evolutionary tracks as magenta lines. All other panels show the histograms of the occurence distribution of 15 different parameters, where the y-axis shows the number of models in a given bin.
              }
         \label{fig:parameter_overview}
\end{figure*}

\subsection{Stellar evolution models}
We used version v-12778 of Modules for Experiments in Stellar Astrophysics \citep[MESA,][]{paxton2011,paxton2013, paxton2015, paxton2018, paxton2019} to calculate the stellar evolution models. The software instrument MESA solves the fully coupled structure and composition equations simultaneously for a one-dimensional, spherically symmetric stellar model \citep{paxton2011}. We refer the reader to the MESA instrument papers for a full description of the numerical methods. More details about the inputs for the microphysics are given in Appendix \ref{app:mesaphysics}.

The physics for the stellar evolution models is similar to the set-up used in the first paper of this series \citep{Steindl2021b}. We used the OPAL opacity tables \citep{seaton2005} and the standard MESA equation of state \citep{paxton2011} while ignoring stellar rotation and magnetic fields. Convection is treated via the mixing length theory developed by \cite{Cox1968}. Convective boundaries are found with the Ledoux criterion and the convective premixing approach \citep{paxton2018} while not allowing for semi-convection. We computed stellar evolution models with mixing lengths $\alpha_{\rm MLT} = 1.8, 2.0, 2.2, 2.4$. Beyond the convective boundary, we employed exponential overshooting \citep{Herwig2000, paxton2011} according to 
\begin{align}
    D_{OV} = D_{\mathrm{conv,}0} \, \exp\left(\frac{-2z}{f H_p}\right)\,,
\end{align}
where $D_{\mathrm{conv,}0}$ is the mixing coefficient near the Schwartzschild boundary, $H_p$ is the pressure scale height, $z$ is the distance to the radiative layer, and $f$ is a free parameter \citep{Herwig2000} set differently for the overshooting towards the surface and the overshooting towards the core of the star. An additional free parameter, $f_0$ describes the position $f_0 H_p$ into the convective zone where the switch between convection and convective overshooting occurs. The resulting four parameters for the overshooting above and below a convective zone in our models are controlled by a parameter $F$ between $0$ and $0.2,$ which results in $f_{\rm above} = F$, $f_{0,{\rm above}} = f_{\rm below} = \frac{F}{2}$, and $f_{0,{\rm below}} =\frac{F}{4}$. The minimum mixing coefficient has values between $D_{\rm mix} = 0$ and $100\, {\rm cm} s^{-1}$.

The initial composition is set to $X_{^1\mathrm{H}} = 1 - X_{^2\mathrm{H}} - X_{^3\mathrm{He}} - X_{^4\mathrm{He}} - Z$, where $X_{^3\mathrm{He}} = 85$~ppm, $X_{^4\mathrm{He}} = Y + 2*Z -  X_{^3\mathrm{He}}$, where $Y = 0.204, 0.214, 0.224,$ or $0.234$ based on the present day surface abundance of the sun and an initial metallicity of $Z,$ which is a value between $0.006$ and $0.024$ spaced out by $0.002$. The mass fractions of metals were taken according to \citet{asplund2009} updated based on \citet{nieva2012} and \citet{Przybilla2013}. The calculation of models is stopped once the central hydrogen fraction has dropped by $0.01$ relatively to the starting value. That is, the evolution includes the earliest part of the main-sequence lifetime. 

The initial models are the same $0.01\,M_\odot$ and $1.5\,R_\odot$ stellar seeds we used in \citet{Steindl2021b}, where we followed the approach of \citet{Kunitomo2017}. Hence, this is similar to the approach of \citet{Stahler1980} and \citet{Hosokawa2011}. These stellar seeds correspond to slightly evolved protostars \citep{Kunitomo2017} with a  high value for the initial entropy of the star \citep{Baraffe2012}. 

For atmospheric boundary conditions, we found that only the temperature-opacity ($T{-}\tau$) relation by \cite{Eddington1926} successfully recreates the instability region for pre-main-sequence $\delta$ Scuti stars \citep{Steindl2021b}. Hence, we used the Eddington atmosphere for our calculations in this work.

\subsubsection{Treatment of accretion}
\label{subsec:treatment_of_accretion}
We describe the energy budget of accretion in Section 5.1.3 of the first paper in this series \citep{Steindl2021b}. Here, we only briefly discuss the details of the treatment of accretion, which are, in part, different to our previous work. 

The energy of the accreted material that is added to the star ($L_{\rm add}$) is controlled by a factor $\beta,$ such that
\begin{align}
    L_{\rm add} = \beta\epsilon \frac{G M \Dot{M}}{R}\, ,
\end{align}
where $\epsilon \leq 0.5$ describes the accretion from a thin disc at the object's equator \citep{Hartmann1997, Baraffe2009}, and $\epsilon = 0.5$ is a common choice \citep{Baraffe2009, Jensen2018, Steindl2021b}. Furthermore, $G$ is the gravitational constant, $M$ is the current mass of the star, $\Dot{M}$ the mass accretion rate, and $R$ the current radius of the star.

For $\beta = 0$, no energy is deposited inside the stellar interior. This case is hence also referred to as cold accretion while $0 < \beta \leq 1$ corresponds to hot accretion. With time-dependent accretion rates, $\beta$ is also expected to vary \citep[e.g.][and references therein]{Baraffe2012, Vorobyov2017, Jensen2018, Elbakyan2019}. Since the accretion rates used in this work are constant, besides an exponential drop towards the end of the accretion phase, we used a fixed value of $\beta$ between $0.05$ and $0.2$.


In MESA, the internal energy of the accreted material is added as extra heat to the star. The distribution of this extra heat in the stellar model is implemented differently in the literature. As a first approach, \citet[][]{Baraffe2010} used a uniform distribution that is most likely too simple to mirror the processes in real stars. As in the first paper of this series, we distributed the heat in an outer region of fractional mass $M_{\rm outer}$ with a linear increase as a function of the mass coordinate $m_r$ \citep[see also][]{Kunitomo2017}  according to
\begin{align}
    l = \frac{L_{\rm add}}{M} {\rm max} \left\{0, \frac{2}{M_{\rm outer}^2} \left(\frac{m_r}{M} -1 + M_{\rm outer} \right)\right\}.
\end{align}

We calculate models with different values of $M_{\rm outer}$, ranging between $0.01$ and $0.2$. Our models are calculated within the constant accretion scenario, meaning that the mass accretion rate $\Dot{M}$ is constant in the beginning of the evolution. In contrast with the first paper of the series, where we used an abrupt change in mass accretion rate, we used an exponential drop with different timescales for the models calculated in this work. Hence, the mass accretion rate is given by
\begin{align}
    \Dot{M} = 
    \begin{cases}
    \Dot{M}_0   & {\rm if} \quad  t < t_{\rm exp}, \\
    \Dot{M}_0 \,\exp\left(\frac{-10(t - t_{\rm exp} )}{\tau}\right) \quad \,\, &{\rm else} \\
    \end{cases}
\end{align}
where $\Dot{M}_0$ is the initial constant accretion rate chosen between  $10^{-7} $ and $10^{-5} \,M_\odot/{\rm yr}$, $t$ is the age of star since the beginning of the simulation, $\tau$ is the timescale for ten e-folding times in the mass-accretion drop chosen between $0.1$ and $1$\, Myr, and $t_{\rm exp}$ is the time until the exponential drop-off calculated as 
\begin{align}
    t_{\rm exp} = \left(M_\star - 0.01- \Dot{M}_0\,\left(1 - \exp(-10)\right)\,\tau \times 10^{6}\,\right)\Dot{M}_0^{-1}.
\end{align}
Here, $M_\star$ is the final mass of the star.




\subsection{Stellar pulsations}
\label{subsec:stellar_pulsations}
The stellar oscillation code GYRE \citep{Townsend2013, Townsend2018, Goldstein2020} uses the equilibrium model resulting from MESA computations and calculates the linear eigenfrequencies for normal oscillation modes. In this work, we used version 6.0.1 of GYRE and refer the reader to the instrument papers \citep{Townsend2013, Townsend2018, Goldstein2020} for numerical details.

We calculated modes with harmonic degrees $l=0, 1,$ and $2$ (radial, dipole, and quadrupole modes) with azimuthal order $m=0$ for p modes for radial orders $n = 1, \cdots, 8$ and the frequency range from $5-100$\,\cd. We chose regularity{-}enforcing inner boundary conditions and outer boundary conditions according to the formulation of \citet{Dziembowski1971}. In contrast to the first paper, we performed the GYRE calculations within the adiabatic framework, since we are only interested in the pulsation frequencies and the calculation of non-adiabatic frequencies is computationally much more expensive.

In general, $\delta$ Scuti stars often show very rich amplitude spectra since the pulsation frequencies of rotating stars are split according to their azimuthal order and rotational properties. In this work, we ignored the splitting of modes and focused on the modes with an azimuthal order of $m=0$. The inclusion of rotational effects will be the subject of future work. 

\subsection{Computational sample}
\label{sec:sample}
To produce a computational sample of pre-main-sequence $\delta$ Scuti stars, we performed a total of $1500$ evolutionary calculations with a random combination of the input physics described above. A summary of the covered parameter space is given in Table \ref{tab:summary_sample}. The convergence of these calculations is dependent on the choice of input physics (see the discussion in Section 5.1.5 of \citet{Steindl2021b}). As a consequence, $1309$ calculations provide pre-main-sequence $\delta$ Scuti models, while the remaining $191$ failed to converge, mostly very early in the evolution. 
During the calculation, we saved a model for the subsequent GYRE calculation if the effective temperature was within the expected instability region ($6 300$\,K - $10 300$\,K) for pre-main-sequence p-mode pulsations \citep{Steindl2021b}. This results in a total of 182\,547 distinct $\delta$ Scuti models. The parameters listed in Table \ref{tab:summary_sample} are direct input parameters for the models. Every model additionally includes a star age, effective temperature, surface gravity, and luminosity. The latter are a direct consequence of the models saved for the GYRE calculations. Figure \ref{fig:parameter_overview} provides an overview over all models calculated for the sample.

 \begin{figure}
   \centering
   \includegraphics[width=\linewidth]{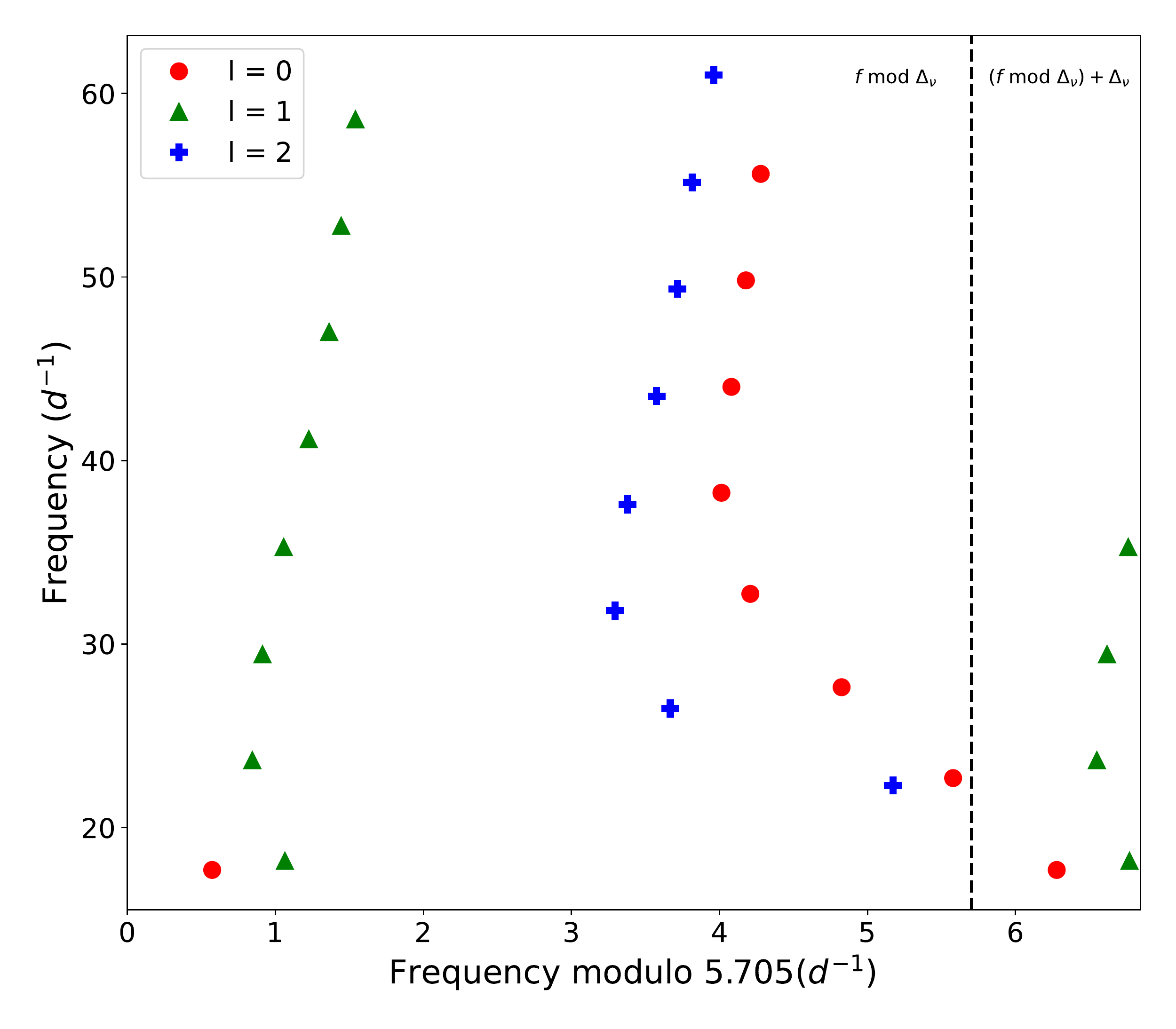}
      \caption{Echelle diagram of theoretical model \#10001. The red, green, and blue symbols correspond to the frequencies of the fundamental and first seven overtones for the radial, dipole, and quadrupole modes, respectively. The echelle diagram is repeated to the right of the dashed black line for better visibility.
              }
         \label{fig:echelle_model10001}
\end{figure}

\section{Modelling approach}
\label{sec:modelling_approach}
In this work, we explored different modelling approaches. For the first ones, we followed previous studies \citep[e.g.][]{Bernabei2009, Chen2019, murphy2021, Kim2021} with the main idea of using a pseudo-$\chi^2$-square statistic:
\begin{align}
\label{xisq}
    \chi^2 = \frac{1}{n}\sum_{i = 1}^n \frac{\left(f_{i, {\rm obs}} - f_{i,{\rm mod}}\right)^2}{\sigma_f},
\end{align}
where $f_{i, {\rm obs}}$ are the observed frequencies, $f_{i,{\rm mod}}$ the model frequencies, and $\sigma_f$ the uncertainty on the observation. We used the Rayleigh limit ($\frac{1}{T}$, where $T$ is the timespan of observations) as the uncertainty of all observed frequencies. In the calculation of the pseudo-$\chi^2$-square statistic, only frequencies with the same mode identification (radial order $n$ and mode degree $l$) are considered, essentially rendering this a modelling of the echelle diagram including the fundamental frequency and the first seven overtones. After the pseudo-$\chi^2$-square statistic was evaluated for all sets of model frequencies, we investigated the properties of all models fulfilling $\chi^2 \leq 1$. Expectation values and uncertainties of stellar parameters are then obtained by additionally applying a spectroscopic observation box (i.e. ignoring all models that do not fall within the $1\sigma$ or $3\sigma$ uncertainty box of $T_{\rm eff}$ and $\log \, g$) and calculating the mean or median as well as the standard deviation of the remaining models. 

A different option is to include the spectroscopic observations within the pseudo-$\chi^2$-square statistic, such that
\begin{align}
\label{spec_xisq}
    \begin{split}
    \chi^2_{\rm spec} = \frac{1}{n + 2}&\left(\sum_{i = 1}^n \frac{\left(f_{i, {\rm obs}} - f_{i,{\rm mod}}\right)^2}{\sigma_f} + \frac{\left(T_{{\rm eff}, {\rm obs}} - T_{{\rm eff}, \rm mod}\right)^2}{\sigma_{T_{\rm eff}}}+ \right.\\&\quad\quad \left.  + \frac{\left(\log\,g_{{\rm obs}} - \log\,g_{ \rm mod}\right)^2}{\sigma_{T_{\log\,g}}} \right).
    \end{split}
\end{align}
However, the latter will reject any model that lies outside the spectroscopic error box if we later apply the condition $\chi^2_{\rm spec} \leq 1$. A more statistically motivated approach is to reject all models with a pseudo-$\chi^2$ statistic greater than the expected $\chi^2$ value for a pre-defined p value, for instance $28.87$ for a p value of $0.05$ and 18 degrees of freedom. 

This modelling approach has significant downsides, as discussed by \citet{Aerts2018} who proposed the use of a Mahalanobis distance for the forward modelling of gravity-mode pulsators. The calculation of the Mahalanobis distance, however, is in need of a variance-covariance matrix obtained from the model grid \citep{Aerts2018}. The model grid calculated in this work is not dense enough to provide a good estimate for the variance-covariance matrix. Hence, we are not able to use the Mahalanobis distance as a model statistic. The work of \cite{Aerts2018} focused on the in-depth modelling of g-mode pulsators for which usual observation time spans of years lead to very low Rayleigh limits. This consequently allows for a much more in-depth study, than the observation time spans of about $30$ to $100$ days usually found for pre-main-sequence $\delta$ Scuti stars. Hence, the modelling of pre-main-sequence $\delta$ Scuti stars pursues lower goals than the forward modelling of gravity mode pulsators. The aim of this work is to investigate whether this simple pseudo-$\chi^2$-square statistic is able to correctly infer the stellar parameters from an identified echelle diagram and to what extent the resulting expectation values and uncertainties can be trusted. As such, we applied five distinct modelling approaches, three of which are based on Equation \ref{xisq}, while the other two are based on Equation \ref{spec_xisq}. The approaches are as follows: (1) all models with $\chi^2 \leq 1$; (2) all models with $\chi^2 \leq 1$ and that lie within the $3\sigma$ spectroscopic error box; (3) all models with $\chi^2 \leq 1$ and that lie within the $1\sigma$ spectroscopic error box; (4) all models with $\chi^2_{\rm spec} \leq 1$; (5) all models with $\chi^2_{\rm spec} \leq \chi^2_{\rm cut}$, where $\chi^2_{\rm cut}$ corresponds to a p value of $0.05$ for the number of free parameters given by the number of frequencies plus two. Approaches 2 and 3 are usually applied in the literature. This study aims to test the validity of these approaches.

\section{Estimating the stellar parameters of theoretical echelle diagrams}
\label{sec:theoretical_echelle_diagrams}

\subsection{Result for a single model}
\label{sec:model10001}

 \begin{figure}
   \centering
   \includegraphics[width=\linewidth]{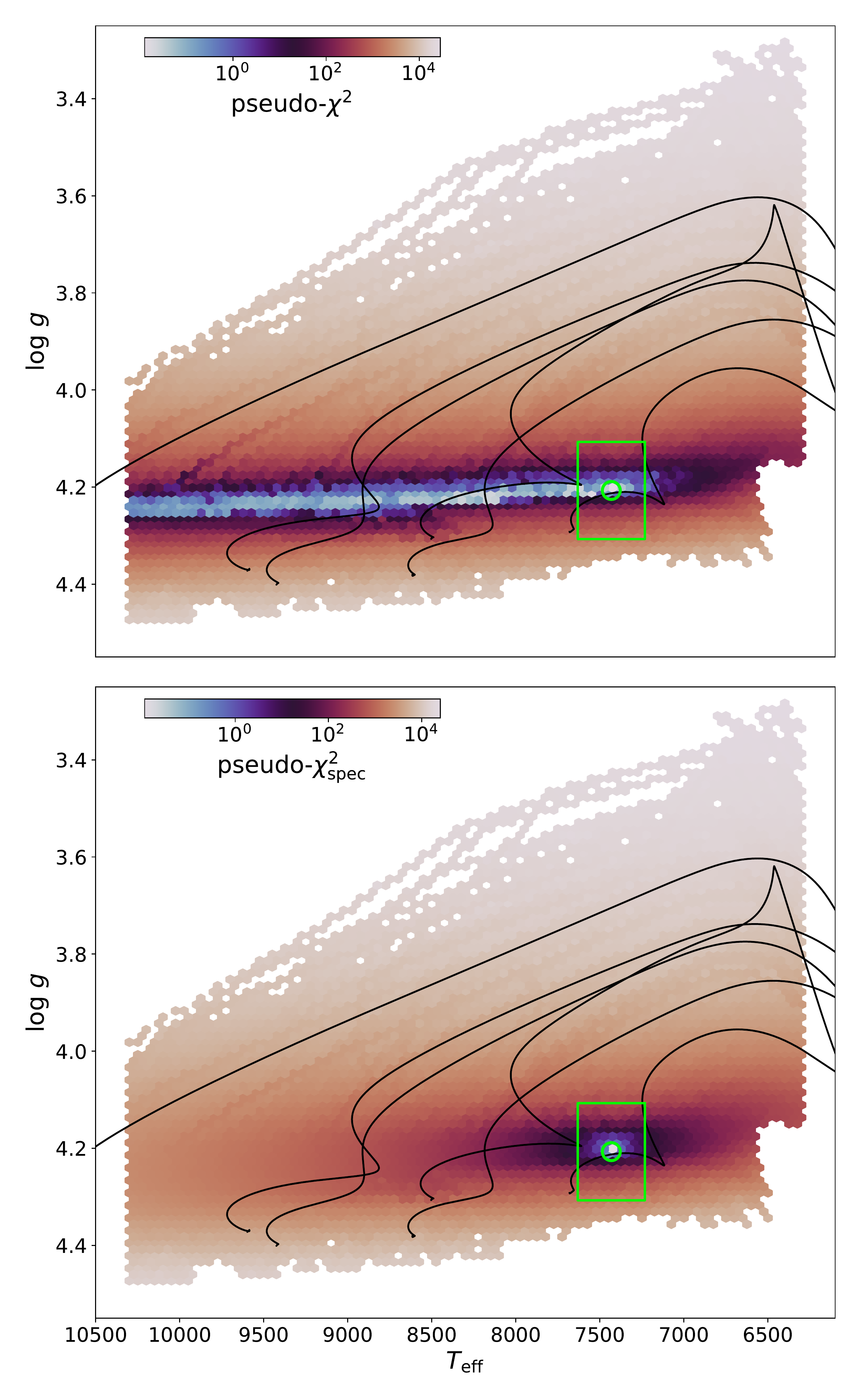}
      \caption{Hexbin plot of the resulting pseudo-$\chi^2$ statistic according to Equation \eqref{xisq} (upper panel) and Equation \eqref{spec_xisq} (lower panel) for the theoretical model \#10001 with a Rayleigh limit of $\frac{1}{27}$~\cd. The colours of the bins show the minimal value of all models that fall within the bin. Black lines show the same evolutionary tracks as in Figure \ref{fig:parameter_overview}. The correct position of model \#10001 is marked with the green circle and the assumed spectroscopic error box is drawn in green. 
              }
         \label{fig:hexbin_model10001}
\end{figure}

 \begin{figure}
   \centering
   \includegraphics[width=\linewidth]{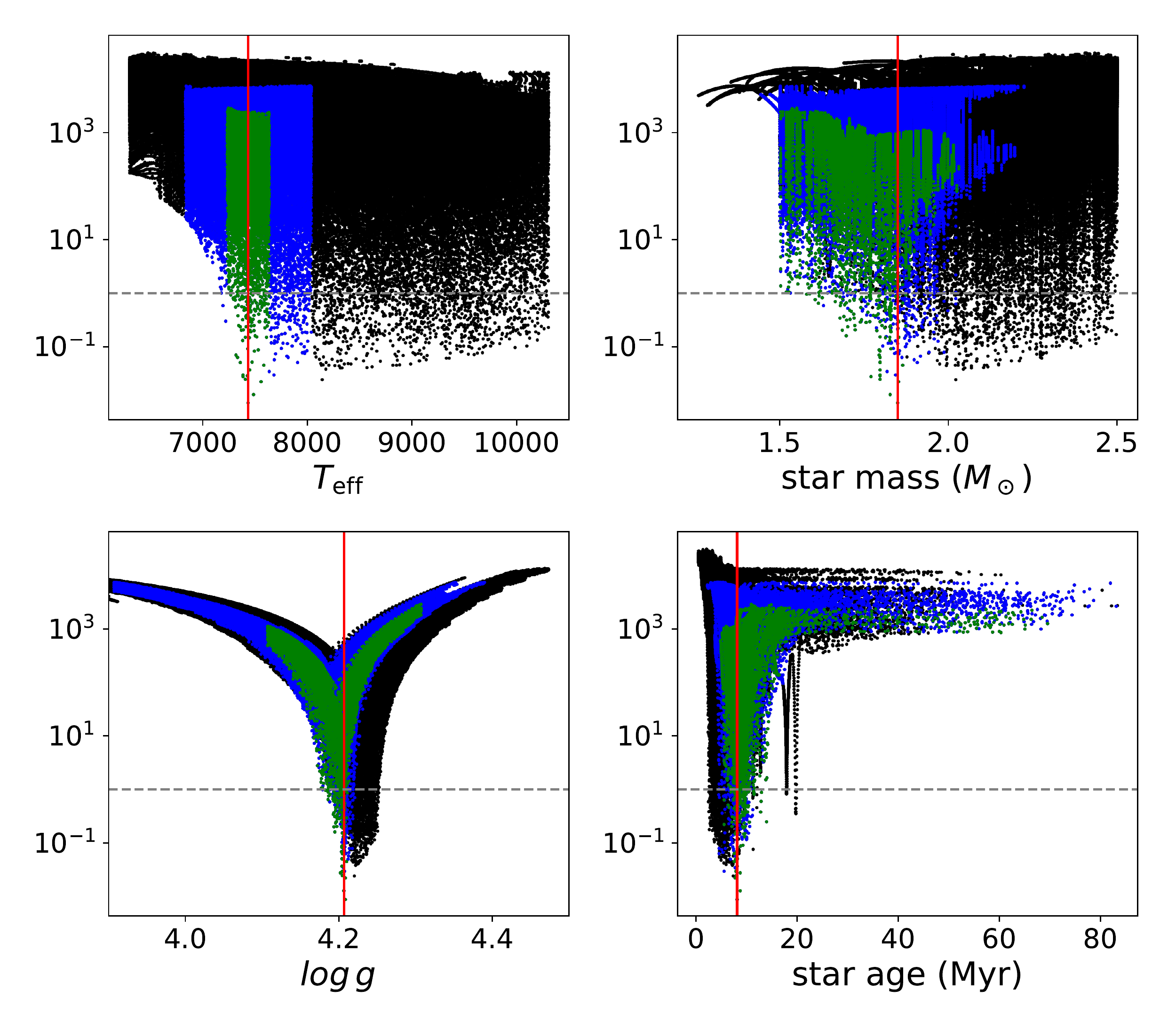}
      \caption{Pseudo-$\chi^2$ statistic distribution of effective temperature, star mass, surface gravity, and age for the frequencies of model \#10001. Models that fall within the $1\sigma$ spectroscopic box are shown in green, those of the $3\sigma$  spectroscopic box are in blue, and all others are in black. The horizontal dashed grey line marks the position of pseudo-$\chi^2 = 1,$ and the vertical red line shows the actual value for model \#10001. 
              }
         \label{fig:dist_model10001}
\end{figure}

We first applied our modelling approaches to one model (\#10001). With an effective temperature of $7432$~K, mass of $1.85\,M_\odot$, and logarithmic surface gravity of $4.208$ at an age of $8.12$~Myr, this model represents typical values of a bona fide pre-main-sequence $\delta$ Scuti star \citep{Zwintz2014, Steindl2020, Steindl2021b}. The echelle spectrum of this model is shown in Figure \ref{fig:echelle_model10001}. To model these theoretical frequencies, we chose a time span of $27$ days, corresponding to one-sector observations of the Transiting Exoplanet Survey Satellite \citep[TESS,][]{Ricker2015}, and perturbed all frequencies with a uniformly chosen random number between $-\frac{1}{T}$ and $+\frac{1}{T}$  where $\frac{1}{T}$ is the resulting Rayleigh limit. In addition, we assumed a typical uncertainty of $200$~K on the effective temperature and $0.1$ on the surface gravity for the spectroscopic error box.

Stellar pulsation modes are represented by spherical harmonics, and higher degree modes suffer from cancellation effects, such that modes with $l \geq 2$ are expected to have lower amplitudes in the light curves than radial and dipole modes \citep{Aerts2010}. In addition, the frequencies of quadrupole modes make a detection of a convincing echelle ridge in observed frequencies much more difficult (see discussion in Section \ref{sec:automatic_echelle_diagrams}).  Hence, for this example we only used the radial and dipole modes of model \#10001. As is evident from the upper panel of Figure \ref{fig:hexbin_model10001}, a range of models with $\log\,g \sim 4.2$ has low values for the pseudo-$\chi^2$ statistics according to Equation \eqref{xisq}. The five models with the lowest pseudo-$\chi^2$ ($\ll 1$) also include, next to model \#10001, model \#52851 with $T_{\rm eff}=8137$~K, $M_\star = 2.02\, M_\odot$, $\log\,g=4.22$, and an age of $7.35$~Myr. The stellar parameters of model \#52851 differ significantly from the the ones of model \#10001. The stellar parameters of many more models, with pseudo-$\chi^2$ statistics well below one, differ to even higher degrees. Thus, it is already evident that the choice of how to enforce the spectroscopic observations plays a major role in the extracted stellar parameters. Figure \ref{fig:dist_model10001} shows the pseudo-$\chi^2$ distribution for these four parameters, with models falling into the spectroscopic error box coloured differently. This further proves that the inclusion of the spectroscopic parameters is a delicate matter. The choice to do so will influence the resulting expectation value and uncertainties. 

As a comparison, the lower panel of Figure \ref{fig:hexbin_model10001} shows the resulting pseudo-$\chi^2_{\rm spec}$ according to Equation \ref{spec_xisq}. The minimum of the pseudo-$\chi^2_{\rm spec}$ is clearly centred around the position of model \#10001 in the Kiel diagram. The latter, however, is expected since it directly influences the  pseudo-$\chi^2_{\rm spec}$ value. To this point, we assumed that the spectroscopic input parameters for the modelling approach were correct. When applied to real data, the spectroscopic input parameters will most likely not exactly be the correct physical values but (hopefully) within the given uncertainty.

To find a preferable method of modelling pre-main sequence $\delta$ Scuti stars, we need to compare their performance in reproducing the stellar parameters and, in an optimal case, as much input physics as possible. For a method to be applicable, it needs to produce the correct parameters even if the spectroscopic parameters are not equivalent to the physical properties of the star. Hence, we also draw the expectation value from a Gaussian distribution around the theoretical model's value with a standard deviation $200$~K.

We applied the five approaches discussed in Section \ref{sec:modelling_approach} to model \#10001. The resulting stellar parameters are given in Table \ref{tab:paramet_model10001_27day}. If we assume the expectation value of the spectroscopic uncertainty box to be correct, only approach 5 exclusively gives correct stellar parameters; that is, the value of the model is within the uncertainty associated with the value extracted by the method. All other approaches disagree for at least one parameter, with approach 2 performing second best and approach 1 performing worst. However, the expectation value of approach 5 is in general not closer to the model parameter, but the associated is uncertainty larger, rendering the results more stable for this model. 

The expectation value of the spectroscopic parameters are most likely not correct in practice. Hence, we investigated the influence of the spectroscopic uncertainty box by drawing a temperature and surface gravity offset from a normal distribution centred around zero with standard deviations of $200$\,K and $0.1$ and applying the methods again. The results, given in Table \ref{tab:paramet_model10001_27day}, draw a similar picture. Methods 2 and 5 perform the best, obtaining the correct value for 86 and 91\% of the stellar parameters. The other approaches fall behind with 29, 63, and 34\% for methods 1, 3, and 5, respectively. We repeated the same analysis for a Rayleigh limit corresponding to a light curve duration of $108$ days (four sectors of TESS observations, which are currently the maximum time base available for a pre-main-sequence $\delta$ Scuti star). The results are shown in Table \ref{tab:paramet_model10001_107d} and are very similar. Again, methods 2 and 5 far outperform methods 1 and 4. Method 3 is the only one to perform better for the assumed $108-$ day precision than the $27-$ day precision. Nevertheless, it trails the performance of methods 2 and 5.

\begin{table}
\begin{footnotesize}
    \caption[]{Extracted stellar parameters of model \#10001 for the different approaches (App.) with a Rayleigh limit of a 27-day light curve. }
    \label{tab:paramet_model10001_27day}
    \tabcolsep=0.04cm
    \begin{tabular*}{\linewidth}{lcccccc}
        \hline
        \noalign{\smallskip}
                & model   &  App. 1  &  App. 2      &  App. 3      &  App. 4    &  App. 5  \\
                & \#10001   & $\chi^2$  & $\chi^2$      & $\chi^2$      & $\chi^2_{\rm spec}$    & $\chi^2_{\rm spec}$  \\
                &           &           & $3\sigma$     &$ 1\sigma$     &        $\leq 1$        &   $\leq 28.87$       \\
        \noalign{\smallskip}
        \hline
        \noalign{\smallskip}
        \multicolumn{7}{l}{spectroscopic parameters assumed correct} \\
        $M_\star$ ($M_\odot$)    &   1.849   &   2.09(20) & \bf{1.83(8)} & 1.78(6)  & 1.77(6)  & \bf{1.75(10)} \\
        $T_{\rm eff}$ (K)    &   7431   &   8800(800) & 7650(210) & \bf{7470(90)}  & \bf{7430(25)}  & \bf{7460(140)} \\
        $\log\,g$    &   4.208   &   4.224(14) & \bf{4.205(6)} & 4.202(5)  & 4.201(6)  & \bf{4.198(11)} \\
        $R$ ($R_\odot$)&   1.772   &  1.85(6) & \bf{1.77(2)} & 1.75(2)  & 1.75(2)  & \bf{1.75(4)} \\
        $L$ ($L_\odot$)&   8.6  &  20(8) & \bf{9.7(1.2)} & \bf{8.5(5)}  & \bf{8.4(2)}  & \bf{8.5(8)} \\
        age (Myr)&   8.12  &  \bf{6.3(2.5)} & \bf{8.4(1.4)} & \bf{8.9(1.3)}  & \bf{8.9(9)}  & \bf{8.8(1.5)} \\
        Z&   0.018 &  \bf{0.018(4)} & \bf{0.018(3)} & \bf{0.018(3)}  & \bf{0.018(3)}  & \bf{0.018(4)} \\
        \noalign{\smallskip}
        \hline
        \multicolumn{7}{l}{spectroscopic parameters assumed $T_{\rm eff} = 7488$\,K and $\log\,g=4.275$} \\
        $M_\star$ ($M_\odot$)    &   1.849   &   2.09(20) & \bf{1.83(8)} & \bf{1.79(6)}  & 1.78(5)  & \bf{1.77(10)} \\
        $T_{\rm eff}$ (K)    &   7431   &   8800(800) & 7700(230) & \bf{7490(110)}  & 7480(30)  & \bf{7500(150)} \\
        $\log\,g$    &   4.208   &   4.224(14) & \bf{4.206(6)} & 4.202(5)  & 4.202(4)  & \bf{4.199(11)} \\
        $R$ ($R_\odot$)&   1.772   &  1.85(6) & \bf{1.77(3)} & 1.75(2)  & 1.75(2)  & \bf{1.75(4)} \\
        $L$ ($L_\odot$)&   8.6  &  20(8) & \bf{10(1.4)} & \bf{8.7(6)}  & \bf{8.6(2)}  & \bf{8.7(9)} \\
        age (Myr)&   8.12  &  \bf{6.3(2.5)} & \bf{8.3(1.5)} & \bf{8.7(1.3)}  & \bf{8.7(9)}  & \bf{8.7(1.4)} \\
        Z&   0.018 &  \bf{0.018(4)} & \bf{0.018(3)} & \bf{0.018(3)}  & \bf{0.018(3)}  & \bf{0.018(4)} \\
        \noalign{\smallskip}
        \hline
        \multicolumn{7}{l}{spectroscopic parameters assumed $T_{\rm eff} = 7553$\,K and $\log\,g=4.127$} \\
        $M_\star$ ($M_\odot$)    &   1.849   &   2.09(20) & \bf{1.84(9)} & \bf{1.80(7)}  & \bf{1.80(5)}  & \bf{1.78(10)} \\
        $T_{\rm eff}$ (K)    &   7431   &   8800(800) & 7740(250) & 7550(110)  & 7560(30)  & \bf{7560(150)} \\
        $\log\,g$    &   4.208   &   4.224(14) & \bf{4.206(6)} & \bf{4.203(6)}  & 4.203(4)  & \bf{4.20(1)} \\
        $R$ ($R_\odot$)&   1.772   &  1.85(6) & \bf{1.77(3)} & 1.76(2)  & 1.76(2)  & \bf{1.75(4)} \\
        $L$ ($L_\odot$)&   8.6  &  20(8) & 10.2(1.5) & \bf{9.0(7)}  & 9.1(2)  & \bf{9.0(9)} \\
        age (Myr)&   8.12  &  \bf{6.3(2.5)} & \bf{8.1(1.5)} & \bf{8.5(1.2)}  & \bf{8.2(1.0)}  & \bf{8.5(1.4)} \\
        Z&   0.018 &  \bf{0.018(4)} & \bf{0.018(3)} & \bf{0.018(3)}  & \bf{0.018(3)}  & \bf{0.018(4)} \\
        \noalign{\smallskip}
        \hline
        \multicolumn{7}{l}{spectroscopic parameters assumed $T_{\rm eff} = 7261$\,K and $\log\,g=4.145$} \\
        $M_\star$ ($M_\odot$)    &   1.849   &   2.09(20) & \bf{1.80(7)} & 1.77(6)  & 1.76(5)  & 1.74(9) \\
        $T_{\rm eff}$ (K)    &   7431   &   8800(800) & \bf{7580(160)} & \bf{7390(60)}  & 7280(30)  & \bf{7330(130)} \\
        $\log\,g$    &   4.208   &   4.224(14) & \bf{4.203(6)} & 4.201(6)  & 4.201(4)  & 4.196(11) \\
        $R$ ($R_\odot$)&   1.772   &  1.85(6) & \bf{1.76(2)} & 1.75(2)  & 1.745(15)  & \bf{1.74(4)} \\
        $L$ ($L_\odot$)&   8.6  &  20(8) & \bf{9.2(9)} & 8.1(3)  & 7.6(2)  & \bf{7.9(7)} \\
        age (Myr)&   8.12  &  \bf{6.3(2.5)} & \bf{8.5(1.5)} & \bf{9.2(1.3)}  & 10.0(1.2)  & \bf{9.2(1.5)} \\
        Z&   0.018 &  \bf{0.018(4)} & \bf{0.018(3)} & \bf{0.018(3)}  & \bf{0.019(2)}  & \bf{0.018(4)} \\
        \noalign{\smallskip}
        \hline
        \multicolumn{7}{l}{spectroscopic parameters assumed $T_{\rm eff} = 7368$\,K and $\log\,g=4.17$} \\
        $M_\star$ ($M_\odot$)    &   1.849   &   2.09(20) & \bf{1.82(7)} & 1.78(6)  & 1.77(5)  & 1.75(9) \\
        $T_{\rm eff}$ (K)    &   7431   &   8800(800) & 7640(200) & \bf{7430(80)}  & 7380(30)  & \bf{7410(140)} \\
        $\log\,g$    &   4.208   &   4.224(14) & \bf{4.204(6)} & 4.202(5)  & 4.201(5)  & \bf{4.197(11)} \\
        $R$ ($R_\odot$)&   1.772   &  1.85(6) & \bf{1.76(2)} & 1.75(2)  & 1.745(18)  & \bf{1.74(4)} \\
        $L$ ($L_\odot$)&   8.6  &  20(8) & \bf{9.6(1.2)} & \bf{8.4(4)}  & 8.1(2)  & \bf{8.2(7)} \\
        age (Myr)&   8.12  &  \bf{6.3(2.5)} & \bf{8.4(1.5)} & \bf{9.0(1.2)}  & 9.4(1.4)  & \bf{9.0(1.5)} \\
        Z&   0.018 &  \bf{0.018(4)} & \bf{0.018(3)} & \bf{0.018(3)}  & \bf{0.02(3)}  & \bf{0.018(4)} \\
        \noalign{\smallskip}
        \hline
        \multicolumn{7}{l}{correct percentage} \\
        \%&   - & 29 & 86 & 63 & 34 & 91 \\
        \noalign{\smallskip}
        \hline

    \end{tabular*}    
    \tablefoot{The expectation value of the extracted parameters adopted as the mean of all chosen models. Only for the metallicity, Z, we adopted the median as the expectation value since the grid is evenly spaced. Values in parentheses give the uncertainties on the extracted parameters adopted from the standard deviation of all models. Parameters that agree with the model are given in bold.
    }
\end{footnotesize}
 \end{table}

\subsection{Result for the model grid}
Approaches 2 and 5 performed best for the modelling of model \#10001. We continued to apply all approaches to more different models. To do so, we drew 75 random echelle diagrams of $l=0$ and $l=1$ modes from our model grid, perturbed the frequencies with a Rayleigh limit corresponding to light curves of $27$ and $108$ days, drew random spectroscopic offsets as we did before, and applied all five modelling approaches. Since we are interested in pre-main-sequence $\delta$ Scuti stars, we limited the age of the models to be below $20$~Myr. Table \ref{tab:all_approaches_overview} presents the results of this exercise as the percentage of stellar parameters that were extracted within $1\sigma$ of the stellar model's value. Table \ref{tab:all_approaches_overview_3sigma} shows the results for  $3\sigma,$ and Table \ref{tab:all_approaches_overview_uncertainty} presents the usual uncertainty for each approach as the median of all 75 models.  Approaches 2 and 5 again perform best, providing the correct stellar parameter within $1\sigma$ in 75\% (76\%) and 74\% (72\%) of the cases for a Rayleigh limit corresponding to a light curve of 27 (108) days. These approaches also yield the correct stellar parameters within $3\sigma$ in almost all cases.  Interestingly, approach 1, which performed worst in the case of model \#10001, does not fall far behind. However, we want to stress that the large amount of models often results in a very large uncertainty on the extracted stellar parameters. Hence, in general we discourage the use of this approach if spectroscopic parameters are available. If no spectroscopic parameters are available, as is often the case for stars in not-well-studied clusters or field stars, this result indicates that meaningful stellar parameters can still be extracted. The reasonable success of approach 3 for the model \#10001 can in general not be reproduced. Only about 50\% of the extracted stellar parameters coincide with the model's parameters. This is, however, not a consequence of the small uncertainties, but rather the stringent spectroscopic constraints being applied. The same argument holds for approach 4, since it requires a $\chi^2_{\rm spec} \leq 1$ and hence an even stronger constraint on the spectroscopic parameters. 

Our results for the model grid verify the results of model \#10001 in the sense that approaches 2 and 5 are to be preferred. The resulting uncertainties (see Table \ref{tab:all_approaches_overview_uncertainty}) are mostly comparable, with the exception that the effective temperature and luminosity are slightly better constrained for approach 5 while the surface gravity and age are slightly better constrained for approach 2. Since the cut-off value for approach 5 is dependent on the number of extracted pulsation frequencies and hence keeps the number of free parameters in mind, this seems to be the most reliable method. However, the implementation of any of these methods is similar enough that we encourage both methods, that is, approaches 2 and 5, be applied. 

\begin{table}
    \caption[]{Percentage rate of extracted stellar parameters that agree with the model's values within $1\sigma$.}
    \label{tab:all_approaches_overview}
    \begin{tabular}{lp{1cm}p{1cm}p{1cm}p{1cm}c}
        \hline
        \noalign{\smallskip}
                   & $\chi^2$  & $\chi^2$      & $\chi^2$      & $\chi^2_{\rm spec}$    & $\chi^2_{\rm spec}$  \\
                   &           & $3\sigma$     &$ 1\sigma$     &        $\leq 1$        &   $\leq 28.87$       \\
                   &     \%      & \%     & \%      &        \%         &   \%        \\
        \noalign{\smallskip}
        \hline
        \noalign{\smallskip}
        \multicolumn{6}{l}{Rayleigh limit corresponding to 27 day light curve} \\
        $M_\star$   & 63 & 76 & 49 & 69 & 76 \\
        $T_{\rm eff}$        & 73 & 83 & 37 & 16 & 64 \\
        $\log\,g$               & 63 & 70 & 46 & 63 & 81 \\
        $R$          & 69 & 71 & 50 & 64 & 81 \\
        $L$          & 74 & 81 & 43 & 31 & 70 \\
        age                & 71 & 77 & 57 & 61 & 84 \\
        Z                       & 64 & 71 & 59 & 71 & 64 \\
        \noalign{\smallskip}
        \hline\noalign{\smallskip}
        total                   & 68 & 75 & 49 & 54 & 74 \\
        \noalign{\smallskip}
        \hline\noalign{\smallskip}
        \multicolumn{6}{l}{Rayleigh limit corresponding to 108 day light curve} \\
        $M_\star$ ($M_\odot$)   & 71 & 77 & 49 & 47 & 74 \\
        $T_{\rm eff}$ (K)       & 71 & 81 & 39 & 13 & 67 \\
        $\log\,g$               & 69 & 67 & 43 & 46 & 79 \\
        $R$ ($R_\odot$)         & 69 & 73 & 44 & 40 & 79 \\
        $L$ ($L_\odot$)         & 74 & 81 & 43 & 16 & 67 \\
        age (Myr)               & 74 & 79 & 46 & 40 & 71 \\
        Z                       & 70 & 77 & 56 & 64 & 64 \\
        \noalign{\smallskip}
        \hline\noalign{\smallskip}
        total                   & 71 & 76 & 48 & 38 & 72 \\
        \noalign{\smallskip}
        \hline

    \end{tabular}  
 \end{table}
 
\subsection{Constraining the input physics of accretion and other free parameters}
The models calculated in this work include free parameters that govern the input physics of accretion. These include the e-folding timescale, the amount of injected heat, the region of injected heat, and the mass-accretion rate. According to our results, echelle diagrams seem to be unresponsive to these parameters; that is, we cannot use the pulsation frequencies to constrain, for example, the mass-accretion rate, since in most cases multiple models with widely different mass accretion rates produce similar pulsation spectra. However, the different input physics of accretion leads to a better estimate of the early evolution of the star in question, and hence contributes to the distribution of stellar models that provide similar theoretical pulsation spectra.

A similar discussion is possible for the other free parameters included in our model grid: the mixing length of convection, envelope mixing, and convective overshoot parameter. Our results again show that, for the Rayleigh limits studied in this work, echelle diagrams are not able to differentiate between the models. However, similarly to the input physics of accretion, these parameters will add to the distribution of models agreeing with the observed frequencies and hence provide a more complete picture of possible evolutionary paths. 
 \begin{figure}
   \centering
   \includegraphics[width=\linewidth]{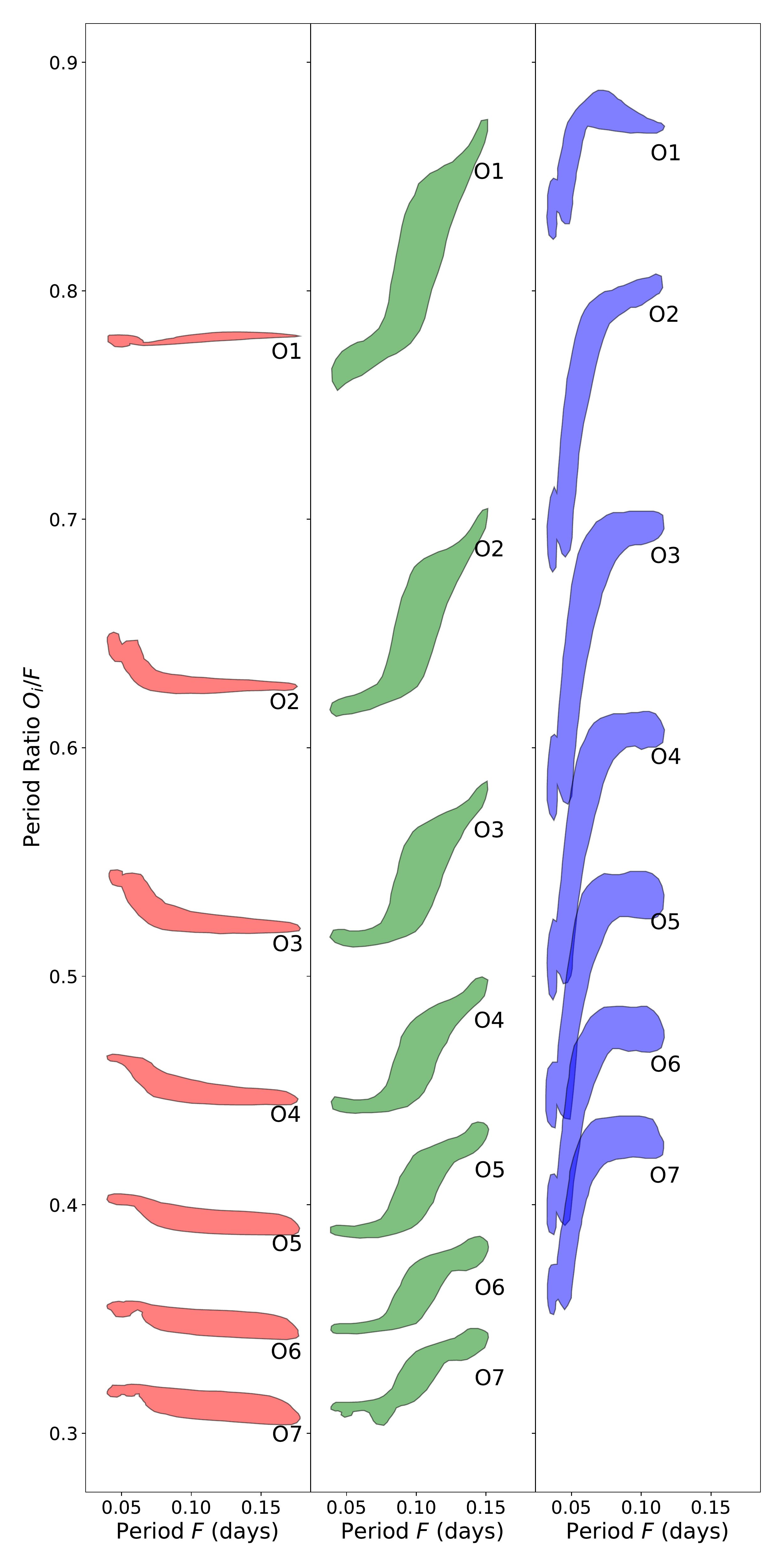}
      \caption{Peterson diagram for the fundamental mode. Left panel: Radial modes ($l=0$). Middle panel: Dipolar modes ($l=1$). Right panel: Quadrupole modes ($l=2$). 
              }
         \label{fig:PetersenF}
\end{figure}

 \begin{figure}
   \centering
   \includegraphics[width=\linewidth]{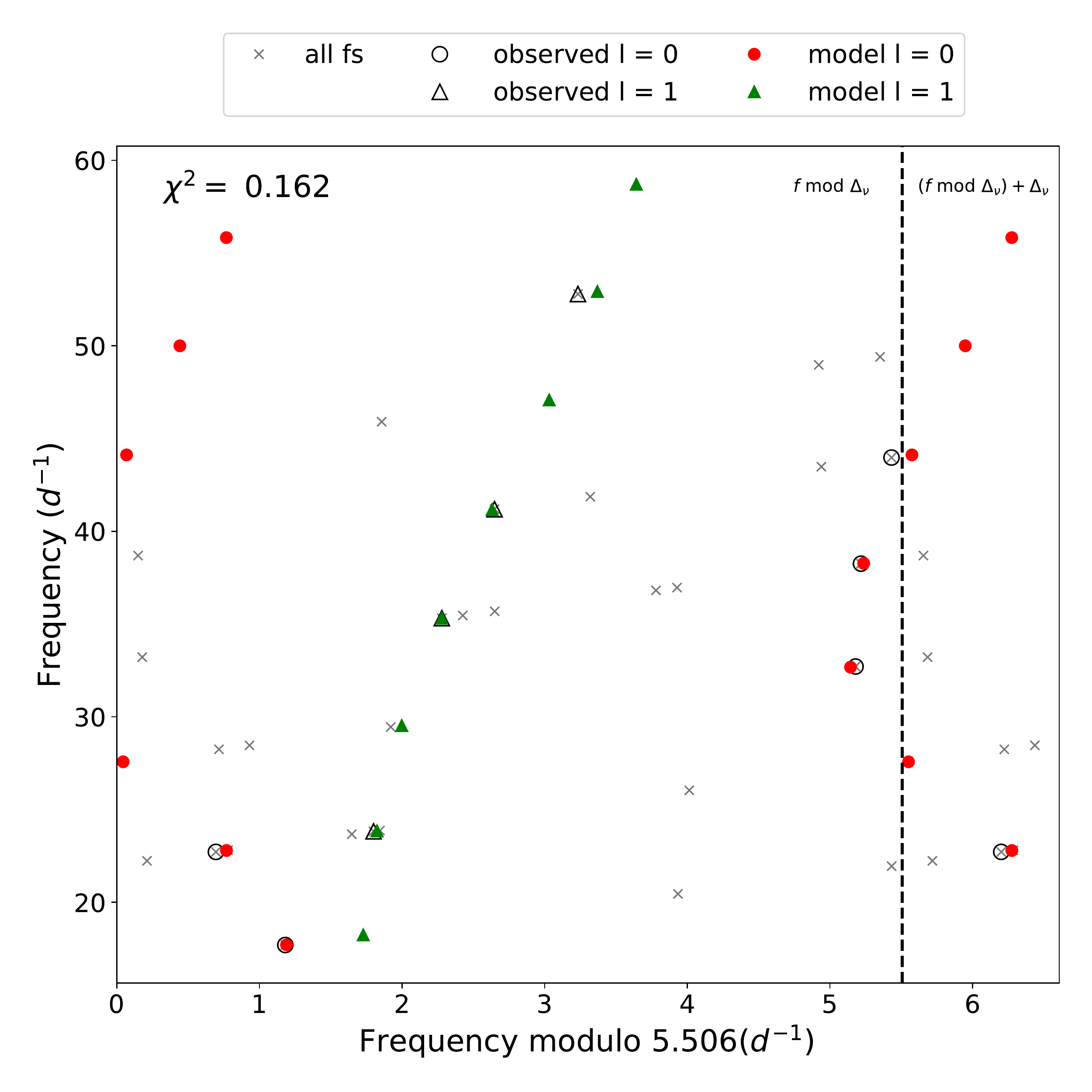}
      \caption{Example output for automatic echelle diagram. The small grey crosses indicate all frequencies given to the algorithm as input. The open circles (triangles) are the radial (dipole) frequencies found by the algorithm, and the coloured circles (triangles) indicate the pulsation modes of one model that agrees with the echelle diagram. Again, the echelle diagram is repeated to the right of the dashed black line for better visibility.
              }
         \label{fig:aed_example}
\end{figure}

\section{Automatic echelle diagrams from observed pulsation frequencies}
\label{sec:automatic_echelle_diagrams}
Amplitude spectra of $\delta$ Scuti stars often show plenty of significant frequencies, such that the extraction of echelle diagrams is mostly impossible. \citet{bedding2020} showed regularities in young main-sequence $\delta$ Scuti stars with echelle diagrams similar to the one shown in Figure \ref{fig:echelle_model10001}. From thousands of $\delta$ Scuti stars observed with TESS and Kepler, they were able to find regularities in the amplitude spectra for only $60$ of them. The dense amplitude spectra of many $\delta$ Scuti stars, most probably due to rotational splitting of non-radial pulsation modes, make it difficult to find regularities in the form of echelle diagrams. Here, we present an automatic approach, based on the period ratios of the pulsation modes. 

It is widely accepted that the ratio between the first overtone and the fundamental mode in radial pulsation modes of $\delta$ Scuti stars show a ratio of $0.77$ \citep{Breger1979}. The ratios of many more combinations of different pulsation modes show similar distinct values, to the extent that diagrams showing such ratios have become known as Peterson diagrams \citep{Petersen1973, Andreasen1983a, Andreasen1983b, Aerts2010}, which show the period ratio of two modes on the y-axis and the period of the pulsation mode with lower radial order on the x-axis. The latter have been used to identify the modes of $\delta$ Scuti stars \citep[i.e.][]{Netzel2021}, and we propose a similar technique to automatically find echelle diagrams of pre-main-sequence $\delta$ Scuti stars. The principle idea is rather simple. We take two significant frequencies and compare their position in the Peterson diagram with the location of period ratios from theoretical models. For the latter, we used all unstable modes in the $\delta$ Scuti instability region from \citet{Steindl2021b}.
Figure \ref{fig:PetersenF} shows the resulting boundaries for the fundamental radial, dipole, and quadrupole modes. We used this to search echelle ridges including the fundamental mode (F) and first seven overtones (O1, O2, O3, O4, O5, O6, O7).  Our approach to finding ridges is based on the following steps.

We start the search for possible ridges by choosing all frequencies that fall within the period range expected for the radial fundamental mode ($0.0425-0.18$\,d). Then, we calculate the period ratios with all other frequencies and check whether the position in the Petersen diagram lies within a boundary of the first seven overtones. If so, the frequency is assigned to the mode of the boundary and added to the ridge. If at least four different frequencies corresponding to different modes are found, we keep the ridge for further analysis. In the case that multiple frequencies have been assigned to the same overtone mode, we split the ridges in such a way that there are two distinct ridges without multiple frequencies assigned to an overtone mode. This procedure is repeated for dipole and quadrupole modes, and for searches starting at the first, second, third, and fourth overtone. This first step results in a list of possible ridges with mode degree and mode identification.

In a second step, we look to verify these ridges by checking every frequency combination within the ridge, whether the resulting period ratio is within the expected boundary in the Petersen diagram. For example, if the ridge contains F, O1, O3, and O4, we check the combinations O1/O3, O1/O4, O3/O4. If more than 70\% (or 85\% for l=2 modes) of the total period ratios fall within the boundary, the ridge is kept for the next step, otherwise removed from the analysis.

Thirdly, we perform a model search by calculating the pseudo-$\chi^2$ value from Equation \ref{xisq} for every model in our model grid. If any model results in $\chi^2 \leq 1$, we keep the ridge. Otherwise, we remove the ridge from the list.

Finally, we combine each ridge with any other ridges with different mode degree and look for possible models explaining both ridges, that is $\chi^2\leq1$ when combining all frequencies of both ridges.

This method finds echelle diagrams from a list of significant frequencies that agree with predictions from pre-main-sequence $\delta$ Scuti models. However, care must be taken when interpreting the results, since it is possible that ridges are built by chance, especially for very dense amplitude spectra. A few other things have to be kept in mind. It is possible that what has been found is not the full ridge; that is, a frequency might be missing because it fell just outside the boundaries in the Petersen diagram. However, by comparing the incomplete ridge with the corresponding model frequencies, it is often simple to identify the missing frequencies. The constraints on the period ratios for degree l=2 modes is less confined than for degree l=0 and l=1 modes (see Figures \ref{fig:PetersenF} and \ref{fig:PetersenO1}, especially). As a consequence, for frequencies with a period of about $0.06$ d, almost all values between $\sim0.35$ and $\sim0.8$ are included within a boundary for $l=2$ modes (see Figure \ref{fig:PetersenF}). Therefore, one finds many more $l=2$ ridges; however, most of them do not correspond to real echelle diagrams. Additionally, the amplitudes of $l=2$ modes are expected to be smaller than radial and dipole modes anyway \citep{Aerts2010}, such that finding an echelle ridge should be less probable. Splitting the frequencies in step 1 can lead to an overwhelming number of different ridges, especially in dense amplitude spectra. In such cases, it is best to only use frequencies with higher significance, that is, a signal-to-noise ratio above six instead of four, as is usual in asteroseismology \citep{breger1993,kuschnig1997}.

Our approach to automatically finding echelle diagrams from the observed frequencies needs to be verified in each case. The simplest approach is to compare any spectroscopic parameters available for the star, with the positions of the models with $\chi^2\leq 1$ in the Kiel diagram. We verified that this approach works by applying it to frequency lists from our model grid. Here, we only show the application to frequencies of model \#10001, but the results were similar for many more models. We applied the algorithm multiple times, each time taking F, O1, O3, O4, and O5 from the radial modes and O1, O2, O3, O4, and O6 from the dipole modes and perturbing them with a Rayleigh limit corresponding to a 27-day light curve. We added twenty randomly chosen frequencies in the range from $20$ to $50$\, \cd. An example output is given in Figure \ref{fig:aed_example}, clearly showing that the algorithm extracted the correct echelle diagram. It remains to be seen how this approach operates for the pulsation frequencies of real stars. We present an example in Section \ref{sec:rediscussion}, but the application to more stars will be the topic of future work.

 \begin{figure}
   \centering
   \includegraphics[width=\linewidth]{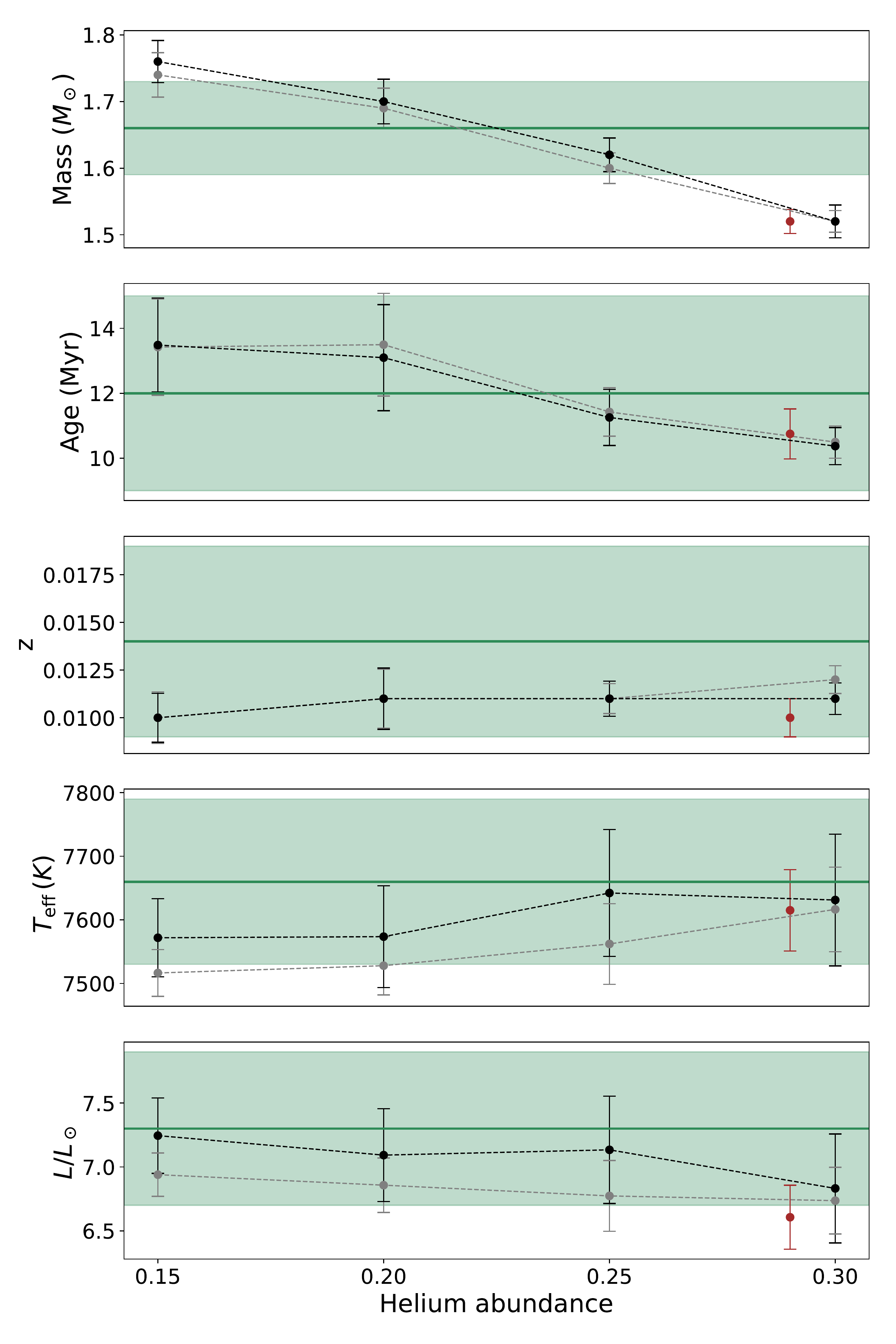}
      \caption{Dependence of stellar parameters of HD 139614 on the choice of initial helium abundance. The black points show the resulting parameters for different values of initial helium abundance when we used the surface gravity as a second classical constraint. The grey counterpart corresponds to the luminosity as second classical constraint. Error bars give the resulting standard deviation. The green area shows our model result, where we adopted the values from approach 5 and the red symbols correspond to the result of \citet{murphy2021}.
              }
         \label{fig:rediscussion_comparison}
\end{figure}

 \begin{figure}
   \centering
   \includegraphics[width=\linewidth]{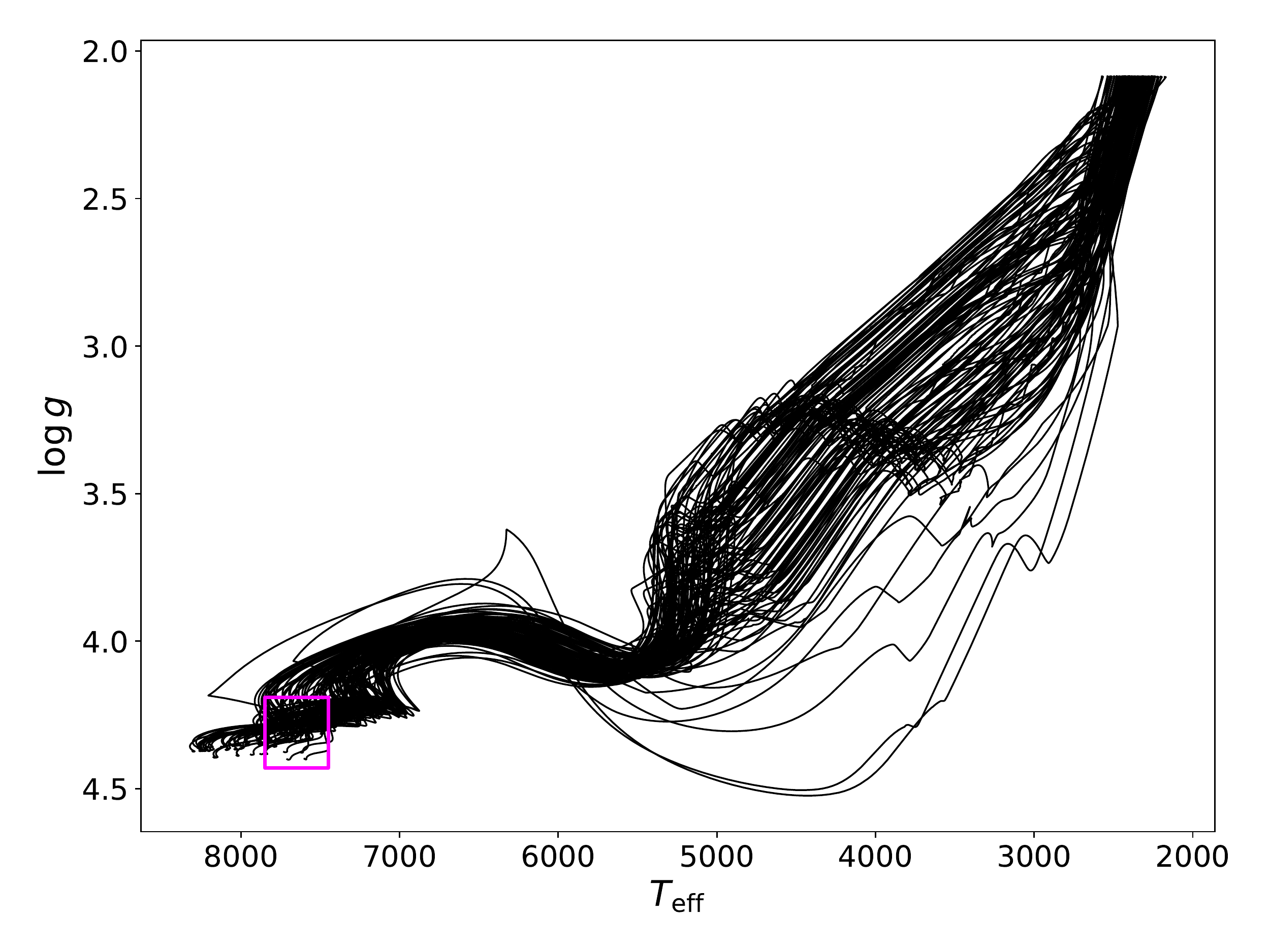}
      \caption{Possible evolutionary history for HD 139614. The figure shows the evolutionary tracks (black lines) of all evolutionary models of the extended grid that provided at least one pulsation model with $\chi^2_{\rm spec} < 22.36$. The $1\sigma$ spectroscopic error box for HD 139614 is shown in magenta. 
              }
         \label{fig:rediscussion_possible_past}
\end{figure}

\section{Re-discussion of the pre-main-sequence $\delta$ Scuti star HD 139614}
\label{sec:rediscussion}
\subsection{Parameter estimation of HD 139614}
\cite{murphy2021} discussed the pre-main-sequence star HD 139614 in detail. They extracted frequencies from the TESS light curve and essentially applied modelling approach 3 to infer the stellar parameters of HD 139614. A major difference between the analysis by \citet{murphy2021} and our analysis presented in the following are the adopted stellar models. \citet{murphy2021} used the classical description of pre-main-sequence models. Their model grid included different values for the initial mass, mixing length, and metallicity. In this work, we employed more advanced models of the pre-main-sequence evolution, starting from accreting protostellar seeds that include many more free parameters (see Section \ref{sec:models}), which leads to a better estimate of the possible evolutionary past of HD 139614. There is also a difference in the adopted value for the initial helium abundance. While \cite{murphy2021} adopted an inital helium abundance of $X_{^4\mathrm{He}} = 0.29$, our values lie in the $X_{^4\mathrm{He}} = 0.216 - 0.282$ range.  An additional difference is that the authors used the stellar luminosity to constrain the spectroscopic parameters, while we used the surface gravity reported by \citet{Fairlamb2015}. Hence, the spectroscopic parameters for HD 139614 used in our study are $\log\,g = 4.31 (12)$\citep{Fairlamb2015} and $T_{\rm eff} = 7650(200)$ \citep{murphy2021}. 

We used the frequencies given by \citet{murphy2021} and our method to automatically find echelle diagrams. We indeed find the same mode identification as \citet{murphy2021} including all other possible identifications mentioned. Our automatic approach additionally yielded their F14 as a fundamental radial mode, a possibility discussed by the authors but ultimately declined as they do not seem to have had corresponding stellar models.  We used the same identification as \citet{murphy2021} for the subsequent modelling. This ensures that differences in resulting stellar parameters are independent of any mode identification ambiguity but are the consequence of an improved stellar modelling. In addition, we later include this fundamental radial mode and discuss the influence on the resulting stellar parameters. 

Since \citet{murphy2021} reported a mass of $1.520(18)\, M_\odot$, we extended our model grid to lower stellar masses. We calculated an additional 300 models, with final masses between $1.2$ and $1.5\,M_\odot$, effectively extending our model grid. Figure \ref{fig:parameter_overview_extended} shows an updated version of Figure \ref{fig:parameter_overview}. We performed the modelling with approaches 2, 3, and 5, since approaches 2 and 5 have proven to be the most reliable, and approach 3 is comparable to that applied by \citet{murphy2021}. 
The evolutionary calculations of our model grid include the earliest parts of the main sequence. Some of these show theoretical pulsation frequencies in agreement with the echelle diagram shown by HD 139614, but they are too old to represent the evolutionary stage of the star. The few too far evolved models will influence the calculation of the mean and standard deviation needed for the extraction of the stellar parameters, with the effect being strongest for the age itself. As such, we decided to apply an age cut-off of 20~Myr\footnote{We experimented with different age cut-offs, looking both at the resulting stellar parameters and the age-$\chi^2$ distribution. While lower cut-offs change the resulting parameters only to a small degree, and within the uncertainties, higher values lead to an asymmetric age-$\chi^2$ distribution and hence to changes in the extracted stellar parameters and their uncertainties. Hence, we believe 20~Myr to be the best choice. In addition, 20 Myr seems to be a reasonable upper limit since HD 139614 is located in the Upper Centaurus Lupus association.} and only use models with ages below that. 

Table \ref{tab:paramet_rediscussion} shows the resulting stellar parameters extracted from the different modelling approaches. As is evident, the values differ significantly from the values reported by \citet{murphy2021}, but they are essentially the same independently of whether or not we include the radial fundamental mode. In addition, it is evident that the extracted uncertainties are larger than the ones reported by \citet{murphy2021}. We obtain the largest difference for the stellar mass, where our modelling yields a value about $0.15\, M_\odot$ larger, depending on the modelling approach. 

\subsection{Model comparison}

We then conducted a comparison between our models and the models presented by \citet{murphy2021} with the aim of illustrating the differences. First, we investigated the effect of using a fixed helium abundance in the models.
For this, we calculated a grid of classical pre-main-sequence models, followed the modelling approach by \citet{murphy2021}, and applied different values for the initial helium abundance.

As is evident in Figure \ref{fig:rediscussion_comparison}, the stellar mass and stellar age are most sensitive to this particular choice of input physics. 
Changing the helium abundance from $Y=0.29$ to $Y=0.25$ leads to a stellar mass for HD 139614 of around $1.62\,M_\odot,$ which is already more than $5\,\sigma$ from  the value reported in \citet{murphy2021}. 

It is important to note that the initial helium abundance is not constrained in the echelle diagram; that is, the models result in a similar goodness of fit. In the same manner, the other input physics (mixing length, e-folding timescale, amount of injected heat, region of injected heat, mass accretion rate, envelope mixing, and overshooting) are not constrained but lead to different stellar evolutions, all producing pulsation frequencies that fit the observed echelle diagram. This result proves that next to the mass and metallicity, the initial helium abundance plays a major role in the determination of stellar parameters. Our model grid additionally allows for variations in envelope mixing and overshooting coefficients. The results of our model grid furthermore indicate that overshooting is an additional important parameter, while the influence of envelope mixing is significantly smaller. In comparison, given the current observational constraints on the pulsation frequencies of HD 139614, the accretion parameters seem to have little importance on the estimated stellar parameters, but they are necessary if additional information about possible previous evolutionary pathways is desired. The inclusion of these parameters allowed us to produce Figure \ref{fig:rediscussion_possible_past}, which shows the possible evolutionary past of HD 139614 in the constant accretion scenario. Evidently, this provides a much more realistic picture than the classical evolutionary concept would let us assume. With HD 139614 still holding onto a protoplanetary disc, this allows for a possible investigation using additional constraints from disc parameters, which shall be the focus of future work. 

To further compare the models applied to HD 139614 by \citet{murphy2021} and in this work, we applied Akaike's information criterion (AIC) and the Bayesian information criterion (BIC). Both are information criteria commonly used for model comparison if the calculation of evidence in the form of likelihood ratios is not possible \citep{Trotta2008, Aerts2018}, as in the case of asteroseismic modelling \citep{Aerts2018}. The two information criteria are defined as
\begin{align*}
    &\text{AIC} = 2k - 2 \ln \mathcal{L}_{max,} \\
    &\text{BIC} = -2\ln \mathcal{L}_{max} + k\ln n\,,
\end{align*}
where $\mathcal{L}_{max}$ is a maximum likelihood estimator, $n$ is the effective number of observations, and $k$ is the number of free parameters in the model. Both are only relevant in comparison, with the basic picture being 'the lower the better' \citep{Aerts2018}. 

Following \citet{Degroote2009}, the AIC and BIC, under the assumption of Gaussian white noise, reduce to
\begin{align*}
    &\text{AIC} =n\ln(\text{RSS}/n) + 2k + n, \\
    &\text{BIC} =n\ln(\text{RSS}/n) + k \ln n\,,
\end{align*}
where RSS is the residual sum of squares, which in our particular case can easily be calculated from the $\chi^2$ value of the best model. We derived the AIC and BIC for our modelling (using the best $\chi^2$ value of $0.107$) and the modelling by \citet{murphy2021}  (using their best $\chi^2$ value of $0.3864$) to be
\begin{align*}
    &\text{AIC}_{\rm this\,work} = -106.9 \quad \quad   \text{AIC}_{\rm Murphy} = -100.5  \quad \quad  \Delta \text{AIC} = -6.4, \\
    &\text{BIC}_{\rm this\,work} = -103.3 \quad \quad  \text{BIC}_{\rm Murphy} = -107.1  \quad \quad  \Delta \text{BIC} = 2.8\,.
\end{align*}
Hence, the AIC provides strong evidence that our model should be preferred, while the BIC provides positive evidence that the analysis of \citet{murphy2021} should be preferred \citep{Aerts2018}. It is important to state that the BIC penalises models with more free parameters stronger than the AIC \citep{Trotta2008}. Additionally, other than, for example, Bayesian evidence, the information criteria apply the penalty to extra parameters regardless of whether they are constrained \citep{Trotta2008}, which is not the case for our ten-dimensional model when applied to HD 139614. As a consequence, the BIC might be overly harsh with regard to our model. 

We investigated this further and calculated an additional model grid, for which we fixed the accretion parameters to these values: accretion rate $\dot{M}_0 = 5\cdot 10^{-6},M_\odot/{\rm yr}$, e-folding time $\tau= 0.5$\, Myr, fractional energy $\beta$= $0.1$, and fractional mass $M_{\rm outer}$ = $0.1$. As discussed above, these parameters are not constrained by the echelle diagram in combination with the observational uncertainty, and hence they impose a strong penalty on the calculation of the AIC and especially the BIC. For this model grid, we calculated a total of 500 stellar evolution models, leading to roughly 50 000 theoretical pulsation spectra. Applying this six-dimensional model grid to HD 139614 with modelling approaches 2, 3, and 5 leads to very similar stellar parameters (see Table \ref{tab:paramet_rediscussion_6dim}) to the ten-dimensional model grid (see Table \ref{tab:paramet_rediscussion}), with only small changes in the expectation values and the uncertainties. The best $\chi^2$ of $0.107$ is also very close to the value obtained from the ten-dimensional model grid. Comparison of the information criteria now results in the following: 
\begin{align*}
    &\text{AIC}_{\rm 6\,d} = -114.9 \quad \quad   \text{AIC}_{\rm Murphy} = -100.5  \quad \quad  \Delta \text{AIC} = -14.4, \\
    &\text{BIC}_{\rm 6\,d} = -117.1 \quad \quad  \text{BIC}_{\rm Murphy} = -107.1  \quad \quad  \Delta \text{BIC} = -10.0\,.
\end{align*}
Hence, the AIC provides very strong evidence and the BIC strong evidence (with $\Delta \text{BIC} =-10$ as the limit between strong and very strong evidence \citep{Aerts2018}) for this model to be preferred over the model by \citet{murphy2021}. 

We illustrate the influence of missing essential ingredients in the model grids used to perform asteroseismic modelling of pre-main-sequence $\delta$ Scuti stars. When a grid consists of  too many fixed parameters, the  resulting fundamental stellar properties will have underestimated uncertainties and possibly inaccurate expectation values. Asteroseismology is a powerful tool, for example, in the description of exoplanets and their host stars. In this regard, it is crucial that the parameters estimated for the host stars be reliable, as any false information will propagate through to the estimated parameters of the planets.

\begin{table}
\begin{footnotesize}
    \caption[]{Extracted stellar parameters of HD 139614 for the different approaches and the two discussed versions of the echelle diagram.}
    \label{tab:paramet_rediscussion}
    \tabcolsep=0.04cm
    \begin{tabular*}{\linewidth}{lcccccc}
        \hline
        \noalign{\smallskip}
                & \citet{murphy2021}    & $\chi^2$      & $\chi^2$      & $\chi^2_{\rm spec}$  \\
                &  modelling               & $3\sigma$     &$ 1\sigma$     &   $\leq 22.36$       \\
        \noalign{\smallskip}
        \hline
        \noalign{\smallskip}
        \multicolumn{7}{l}{same mode identification as in \citet{murphy2021}} \\
        $M_\star$ ($M_\odot$)    &   1.520(18)   &  1.7(1) & 1.66(6)  & 1.66(7)   \\
        $T_{\rm eff}$ (K)    &   7615(64)   &   7830(250) & 7670(110) & 7650(130)  \\
        $\log\,g$    &   -   &   4.288(8) & 4.285(5) & 4.284(8)\\
        $R$ ($R_\odot$)&   -   &  1.55(3) & 1.53(2) & 1.54(3)  \\
        $L$ ($L_\odot$)&   6.7(3)  &  8.2(1.3) & 7.4(5)  & 7.3(6) \\
        age (Myr)&   10.75(77)  &  12(3) & 12(3) & 12(3)  \\
        Z&   0.010(1) &  0.014(5) & 0.012(4) & 0.014(5)\\
        \noalign{\smallskip}
        \hline
        \noalign{\smallskip}
                & \citet{murphy2021}    & $\chi^2$      & $\chi^2$      & $\chi^2_{\rm spec}$  \\
                &  modelling               & $3\sigma$     &$ 1\sigma$     &   $\leq 23.68$       \\
        \noalign{\smallskip}
        \hline
        \noalign{\smallskip}
        \multicolumn{7}{l}{additionally include the fundamental radial mode} \\
        $M_\star$ ($M_\odot$)    &   1.520(18)   &  1.71(9) & 1.66(6)  & 1.66(7)   \\
        $T_{\rm eff}$ (K)    &   7615(64)   &   7830(250) & 7680(110) & 7660(130)  \\
        $\log\,g$    &   -   &   4.289(7) & 4.286(5) & 4.284(8)\\
        $R$ ($R_\odot$)&   -   &  1.55(3) & 1.54(2) & 1.54(3)  \\
        $L$ ($L_\odot$)&   6.7(3)  &  8.2(1.3) & 7.4(5)  & 7.3(6) \\
        age (Myr)&   10.75(77)  &  12(3) & 12(3) & 12(3))  \\
        Z&   0.010(1) &  0.014(4) & 0.012(3) & 0.014(5)\\
        \noalign{\smallskip}
        \hline

    \end{tabular*}    
    \tablefoot{The expectation value of the extracted parameter adopted as the mean of all chosen models. Only for the metallicity, Z, did we adopt the median as the expectation value since the grid is evenly spaced. Values in parentheses give the uncertainty on the extracted parameter adopted from the standard deviation of all models.
    }
\end{footnotesize}
 \end{table}

\section{Conclusion}
\label{sec:conlusion}
In this work, we investigated different approaches in the pulsational modelling of pre-main-sequence $\delta$ Scuti stars. For this, we calculated a grid of stellar evolution models starting from the protostellar accretion phase, including various free parameters such as the final mass, initial composition, mixing parameter, and details of the accretion process. The resulting equilibrium models of pre-main-sequence stars are used as input for adiabatic stellar pulsation calculations to deliver a grid of pulsation spectra. 
We introduced a total of five different modelling approaches all based on a pseudo-$\chi^2$ statistic as a goodness of fit indicator for the pulsation frequencies. While the first approach uses a simple pseudo-$\chi^2$ statistic, the other approaches additionally make use of spectroscopic constraints. Approaches 2 and 3 only consider models that fall within a given spectroscopic error box, while approaches 4 and 5 include the spectroscopic constraints within the pseudo-$\chi^2$ statistic itself. All different modelling techniques are first applied to a single theoretical model and then to a set of 75 randomly chosen models from our model grid. During this endeavour, we include frequency shifts according to a Rayleigh limit corresponding to light curves with time bases of 27 and 108 days. In addition, we draw a random offset for the the expectation value of the spectroscopic parameters from a Gaussian distribution to insure the inclusion of measurement uncertainties in these aspects as well. 

Our results clearly indicate that two approaches (i.e. 2 and 5) are to be preferred in the modelling of pre-main-sequence $\delta$ Scuti stars. Other approaches often result in unreliable stellar parameters (approaches 3 and 4) or in very large uncertainties (approach 1). According to our results, we conclude that modelling approach 5, which included the spectroscopic parameters in the calculation of the pseudo-$\chi^2$ statistic and sets the pseudo-$\chi^2$ cut-off depending on the amount of free parameters, is the most reliable. We note that approach 2, which uses the simpler pseudo-$\chi^2$ statistics and only considers models within $3\sigma$ of the spectroscopic constraints, does not perform worse. The work presented here demonstrates that typical uncertainties for the stellar radius are in the range of $0.05\,R_\odot$. For lots of stars, we expect resulting uncertainties to be smaller than this since this is the median value within all studied models.

In addition, we introduced a simple algorithm for the extraction of echelle diagrams in the observed pulsation spectra of pre-main-sequence $\delta$ Scuti stars. This algorithm is based on theoretical Petersen diagrams taken from the non adiabatic pre-main-sequence models of \citet{Steindl2021b}. We show that the algorithm correctly identifies the echelle ridge for perturbed frequencies of our modelling grid.

Finally, we combine the presented work and apply it to rediscuss the asteroseismic modelling of the pre-main-sequence $\delta$ Scuti star HD 139614. This star was recently modelled by \citet{murphy2021}, but the applied model grid used an outdated description of pre-main-sequence models and a less dimensional model grid. Our results indicate that this simplified model grid is insufficient to properly describe the complete picture of the evolutionary pathways of HD 139614. Hence, our study results in different stellar parameters, which most notably have higher uncertainties. At this point, it is important to stress that these uncertainties stem from an improved grid of stellar models and an improved modelling approach and should hence depict the current limitations of asteroseismic modelling of pre-main-sequence stars. Additionally, we repeated the analysis of \citet{murphy2021} with classical model grids to show the dependence of the results on the initial helium composition. Most notably, this explains the difference in extracted stellar mass for HD 139614 and results in far greater uncertainty with regard to stellar age than mentioned by \citet{murphy2021}.We performed a model comparison by applying commonly used information criteria to illustrate the differences between our approach and the one by \citet{murphy2021}.

This work clearly shows that producing an extensive grid of stellar models to perform this kind of asteroseismic modelling of pre-main-sequence $\delta$ Scuti stars is unavoidable. If, in contrast, simpler model grids are utilised, resulting parameters are often extracted with incorrect expectation values and underestimated uncertainties.

\begin{acknowledgements}
    We are grateful to Bill Paxton and his collaborators for their valuable work on the stellar evolution code MESA. 
    This research has made use of matplotlib, a Python library for publication quality graphics \citep{Hunter:2007}; SciPy \citep{Virtanen_2020}; NumPy \citep{van2011numpy}; MESA SDK for Linux (Version 20.3.1) \citep{townsend2020}.
\end{acknowledgements}

\bibliographystyle{aa} 
\bibliography{bib} 

\begin{thebibliography}{75}
\expandafter\ifx\csname natexlab\endcsname\relax\def\natexlab#1{#1}\fi

\bibitem[{{Aerts} {et~al.}(2010){Aerts}, {Christensen-Dalsgaard}, \&
  {Kurtz}}]{Aerts2010}
{Aerts}, C., {Christensen-Dalsgaard}, J., \& {Kurtz}, D.~W. 2010,
  {Asteroseismology}

\bibitem[{{Aerts} {et~al.}(2018){Aerts}, {Molenberghs}, {Michielsen},
  {Pedersen}, {Bj{\"o}rklund}, {Johnston}, {Mombarg}, {Bowman}, {Buysschaert},
  {P{\'a}pics}, {Sekaran}, {Sundqvist}, {Tkachenko}, {Truyaert}, {Van Reeth},
  \& {Vermeyen}}]{Aerts2018}
{Aerts}, C., {Molenberghs}, G., {Michielsen}, M., {et~al.} 2018, \apjs, 237, 15

\bibitem[{{Andreasen}(1983)}]{Andreasen1983b}
{Andreasen}, G.~K. 1983, \aap, 121, 250

\bibitem[{{Andreasen} {et~al.}(1983){Andreasen}, {Hejlesen}, \&
  {Petersen}}]{Andreasen1983a}
{Andreasen}, G.~K., {Hejlesen}, P.~M., \& {Petersen}, J.~O. 1983, \aap, 121,
  241

\bibitem[{{Angulo} {et~al.}(1999){Angulo}, {Arnould}, {Rayet}, {Descouvemont},
  {Baye}, {Leclercq-Willain}, {Coc}, {Barhoumi}, {Aguer}, {Rolfs}, {Kunz},
  {Hammer}, {Mayer}, {Paradellis}, {Kossionides}, {Chronidou}, {Spyrou},
  {degl'Innocenti}, {Fiorentini}, {Ricci}, {Zavatarelli}, {Providencia},
  {Wolters}, {Soares}, {Grama}, {Rahighi}, {Shotter}, \& {Lamehi
  Rachti}}]{Angulo1999}
{Angulo}, C., {Arnould}, M., {Rayet}, M., {et~al.} 1999, Nuclear Physics A,
  656, 3

\bibitem[{{Asplund} {et~al.}(2009){Asplund}, {Grevesse}, {Sauval}, \&
  {Scott}}]{asplund2009}
{Asplund}, M., {Grevesse}, N., {Sauval}, A.~J., \& {Scott}, P. 2009, \araa, 47,
  481

\bibitem[{{Baraffe} \& {Chabrier}(2010)}]{Baraffe2010}
{Baraffe}, I. \& {Chabrier}, G. 2010, \aap, 521, A44

\bibitem[{{Baraffe} {et~al.}(2009){Baraffe}, {Chabrier}, \&
  {Gallardo}}]{Baraffe2009}
{Baraffe}, I., {Chabrier}, G., \& {Gallardo}, J. 2009, \apjl, 702, L27

\bibitem[{{Baraffe} {et~al.}(2012){Baraffe}, {Vorobyov}, \&
  {Chabrier}}]{Baraffe2012}
{Baraffe}, I., {Vorobyov}, E., \& {Chabrier}, G. 2012, \apj, 756, 118

\bibitem[{{Bedding} {et~al.}(2020){Bedding}, {Murphy}, {Hey}, {Huber}, {Li},
  {Smalley}, {Stello}, {White}, {Ball}, {Chaplin}, {Colman}, {Fuller},
  {Gaidos}, {Harbeck}, {Hermes}, {Holdsworth}, {Li}, {Li}, {Mann}, {Reese},
  {Sekaran}, {Yu}, {Antoci}, {Bergmann}, {Brown}, {Howard}, {Ireland},
  {Isaacson}, {Jenkins}, {Kjeldsen}, {McCully}, {Rabus}, {Rains}, {Ricker},
  {Tinney}, \& {Vanderspek}}]{bedding2020}
{Bedding}, T.~R., {Murphy}, S.~J., {Hey}, D.~R., {et~al.} 2020, \nat, 581, 147

\bibitem[{{Bernabei} {et~al.}(2009){Bernabei}, {Ripepi}, {Ruoppo}, {Marconi},
  {Monteiro}, {Rodriguez}, {Oswalt}, {Leccia}, {Palla}, {Catanzaro}, {Amado},
  {Lopez-Gonzalez}, \& {Aceituno}}]{Bernabei2009}
{Bernabei}, S., {Ripepi}, V., {Ruoppo}, A., {et~al.} 2009, \aap, 501, 279

\bibitem[{{Breger}(1979)}]{Breger1979}
{Breger}, M. 1979, \pasp, 91, 5

\bibitem[{{Breger} {et~al.}(1993){Breger}, {Stich}, {Garrido}, {Martin},
  {Jiang}, {Li}, {Hube}, {Ostermann}, {Paparo}, \& {Scheck}}]{breger1993}
{Breger}, M., {Stich}, J., {Garrido}, R., {et~al.} 1993, \aap, 271, 482

\bibitem[{{Buchler} \& {Yueh}(1976)}]{Buchler1976}
{Buchler}, J.~R. \& {Yueh}, W.~R. 1976, \apj, 210, 440

\bibitem[{{Cassisi} {et~al.}(2007){Cassisi}, {Potekhin}, {Pietrinferni},
  {Catelan}, \& {Salaris}}]{Cassisi2007}
{Cassisi}, S., {Potekhin}, A.~Y., {Pietrinferni}, A., {Catelan}, M., \&
  {Salaris}, M. 2007, \apj, 661, 1094

\bibitem[{{Chen} \& {Li}(2019)}]{Chen2019}
{Chen}, X. \& {Li}, Y. 2019, \apj, 872, 156

\bibitem[{{Chugunov} {et~al.}(2007){Chugunov}, {Dewitt}, \&
  {Yakovlev}}]{Chugunov2007}
{Chugunov}, A.~I., {Dewitt}, H.~E., \& {Yakovlev}, D.~G. 2007, \prd, 76, 025028

\bibitem[{{Cox} \& {Giuli}(1968)}]{Cox1968}
{Cox}, J.~P. \& {Giuli}, R.~T. 1968, {Principles of stellar structure}

\bibitem[{{Cyburt} {et~al.}(2010){Cyburt}, {Amthor}, {Ferguson}, {Meisel},
  {Smith}, {Warren}, {Heger}, {Hoffman}, {Rauscher}, {Sakharuk}, {Schatz},
  {Thielemann}, \& {Wiescher}}]{Cyburt2010}
{Cyburt}, R.~H., {Amthor}, A.~M., {Ferguson}, R., {et~al.} 2010, \apjs, 189,
  240

\bibitem[{{Degroote} {et~al.}(2009){Degroote}, {Briquet}, {Catala},
  {Uytterhoeven}, {Lefever}, {Morel}, {Aerts}, {Carrier}, {Auvergne}, {Baglin},
  \& {Michel}}]{Degroote2009}
{Degroote}, P., {Briquet}, M., {Catala}, C., {et~al.} 2009, \aap, 506, 111

\bibitem[{{Dupret} {et~al.}(2005){Dupret}, {Grigahc{\`e}ne}, {Garrido},
  {Gabriel}, \& {Scuflaire}}]{Dupret2005}
{Dupret}, M.~A., {Grigahc{\`e}ne}, A., {Garrido}, R., {Gabriel}, M., \&
  {Scuflaire}, R. 2005, \aap, 435, 927

\bibitem[{{Dziembowski}(1971)}]{Dziembowski1971}
{Dziembowski}, W.~A. 1971, \actaa, 21, 289

\bibitem[{{Eddington}(1926)}]{Eddington1926}
{Eddington}, A.~S. 1926, {The Internal Constitution of the Stars}

\bibitem[{{Elbakyan} {et~al.}(2019){Elbakyan}, {Vorobyov}, {Rab}, {Meyer},
  {G{\"u}del}, {Hosokawa}, \& {Yorke}}]{Elbakyan2019}
{Elbakyan}, V.~G., {Vorobyov}, E.~I., {Rab}, C., {et~al.} 2019, \mnras, 484,
  146

\bibitem[{{Fairlamb} {et~al.}(2015){Fairlamb}, {Oudmaijer}, {Mendigut{\'\i}a},
  {Ilee}, \& {van den Ancker}}]{Fairlamb2015}
{Fairlamb}, J.~R., {Oudmaijer}, R.~D., {Mendigut{\'\i}a}, I., {Ilee}, J.~D., \&
  {van den Ancker}, M.~E. 2015, \mnras, 453, 976

\bibitem[{{Ferguson} {et~al.}(2005){Ferguson}, {Alexander}, {Allard}, {Barman},
  {Bodnarik}, {Hauschildt}, {Heffner-Wong}, \& {Tamanai}}]{Ferguson2005}
{Ferguson}, J.~W., {Alexander}, D.~R., {Allard}, F., {et~al.} 2005, \apj, 623,
  585

\bibitem[{{Fuller} {et~al.}(1985){Fuller}, {Fowler}, \& {Newman}}]{Fuller1985}
{Fuller}, G.~M., {Fowler}, W.~A., \& {Newman}, M.~J. 1985, \apj, 293, 1

\bibitem[{{Goldstein} \& {Townsend}(2020)}]{Goldstein2020}
{Goldstein}, J. \& {Townsend}, R.~H.~D. 2020, \apj, 899, 116

\bibitem[{{Gruber} {et~al.}(2012){Gruber}, {Saio}, {Kuschnig}, {Fossati}, {Hand
  ler}, {Zwintz}, {Weiss}, {Matthews}, {Guenther}, {Moffat}, {Rucinski}, \&
  {Sasselov}}]{Gruber2012}
{Gruber}, D., {Saio}, H., {Kuschnig}, R., {et~al.} 2012, \mnras, 420, 291

\bibitem[{{Hartmann} {et~al.}(1997){Hartmann}, {Cassen}, \&
  {Kenyon}}]{Hartmann1997}
{Hartmann}, L., {Cassen}, P., \& {Kenyon}, S.~J. 1997, \apj, 475, 770

\bibitem[{{Hayashi}(1961)}]{Hayashi1961}
{Hayashi}, C. 1961, \pasj, 13, 450

\bibitem[{{Henyey} {et~al.}(1955){Henyey}, {Lelevier}, \&
  {Lev{\'e}e}}]{Henyey1955}
{Henyey}, L.~G., {Lelevier}, R., \& {Lev{\'e}e}, R.~D. 1955, \pasp, 67, 154

\bibitem[{{Herwig}(2000)}]{Herwig2000}
{Herwig}, F. 2000, \aap, 360, 952

\bibitem[{{Hosokawa} {et~al.}(2011){Hosokawa}, {Offner}, \&
  {Krumholz}}]{Hosokawa2011}
{Hosokawa}, T., {Offner}, S. S.~R., \& {Krumholz}, M.~R. 2011, \apj, 738, 140

\bibitem[{Hunter(2007)}]{Hunter:2007}
Hunter, J.~D. 2007, Computing In Science \& Engineering, 9, 90

\bibitem[{{Iben}(1965)}]{Iben1965}
{Iben}, Icko, J. 1965, \apj, 141, 993

\bibitem[{{Iglesias} \& {Rogers}(1993)}]{Iglesias1993}
{Iglesias}, C.~A. \& {Rogers}, F.~J. 1993, \apj, 412, 752

\bibitem[{{Iglesias} \& {Rogers}(1996)}]{Iglesias1996}
{Iglesias}, C.~A. \& {Rogers}, F.~J. 1996, \apj, 464, 943

\bibitem[{{Itoh} {et~al.}(1996){Itoh}, {Hayashi}, {Nishikawa}, \&
  {Kohyama}}]{Itoh1996}
{Itoh}, N., {Hayashi}, H., {Nishikawa}, A., \& {Kohyama}, Y. 1996, \apjs, 102,
  411

\bibitem[{{Jensen} \& {Haugb{\o}lle}(2018)}]{Jensen2018}
{Jensen}, S.~S. \& {Haugb{\o}lle}, T. 2018, \mnras, 474, 1176

\bibitem[{{Kim} {et~al.}(2021){Kim}, {Lee}, {Lee}, {Lee}, {Lee}, {Hong}, {Cha},
  {Kim}, \& {Park}}]{Kim2021}
{Kim}, S.-L., {Lee}, J.~W., {Lee}, C.-U., {et~al.} 2021, \aj, 162, 212

\bibitem[{{Kunitomo} {et~al.}(2017){Kunitomo}, {Guillot}, {Takeuchi}, \&
  {Ida}}]{Kunitomo2017}
{Kunitomo}, M., {Guillot}, T., {Takeuchi}, T., \& {Ida}, S. 2017, \aap, 599,
  A49

\bibitem[{{Kuschnig} {et~al.}(1997){Kuschnig}, {Weiss}, {Gruber}, {Bely}, \&
  {Jenkner}}]{kuschnig1997}
{Kuschnig}, R., {Weiss}, W.~W., {Gruber}, R., {Bely}, P.~Y., \& {Jenkner}, H.
  1997, \aap, 328, 544

\bibitem[{{Langanke} \& {Mart{\'{\i}}nez-Pinedo}(2000)}]{Langanke2000}
{Langanke}, K. \& {Mart{\'{\i}}nez-Pinedo}, G. 2000, Nuclear Physics A, 673,
  481

\bibitem[{{Larson}(1969)}]{Larson1969}
{Larson}, R.~B. 1969, \mnras, 145, 271

\bibitem[{{Masunaga} \& {Inutsuka}(2000)}]{Masunaga2000}
{Masunaga}, H. \& {Inutsuka}, S.-i. 2000, \apj, 531, 350

\bibitem[{{Murphy} {et~al.}(2021){Murphy}, {Joyce}, {Bedding}, {White}, \&
  {Kama}}]{murphy2021}
{Murphy}, S.~J., {Joyce}, M., {Bedding}, T.~R., {White}, T.~R., \& {Kama}, M.
  2021, \mnras, 502, 1633

\bibitem[{{Netzel} {et~al.}(2022){Netzel}, {Pietrukowicz}, {Soszy{\'n}ski}, \&
  {Wrona}}]{Netzel2021}
{Netzel}, H., {Pietrukowicz}, P., {Soszy{\'n}ski}, I., \& {Wrona}, M. 2022,
  \mnras, 510, 1748

\bibitem[{{Nieva} \& {Przybilla}(2012)}]{nieva2012}
{Nieva}, M.~F. \& {Przybilla}, N. 2012, \aap, 539, A143

\bibitem[{{Oda} {et~al.}(1994){Oda}, {Hino}, {Muto}, {Takahara}, \&
  {Sato}}]{Oda1994}
{Oda}, T., {Hino}, M., {Muto}, K., {Takahara}, M., \& {Sato}, K. 1994, Atomic
  Data and Nuclear Data Tables, 56, 231

\bibitem[{{Paxton} {et~al.}(2011){Paxton}, {Bildsten}, {Dotter}, {Herwig},
  {Lesaffre}, \& {Timmes}}]{paxton2011}
{Paxton}, B., {Bildsten}, L., {Dotter}, A., {et~al.} 2011, \apjs, 192, 3

\bibitem[{{Paxton} {et~al.}(2013){Paxton}, {Cantiello}, {Arras}, {Bildsten},
  {Brown}, {Dotter}, {Mankovich}, {Montgomery}, {Stello}, {Timmes}, \&
  {Townsend}}]{paxton2013}
{Paxton}, B., {Cantiello}, M., {Arras}, P., {et~al.} 2013, \apjs, 208, 4

\bibitem[{{Paxton} {et~al.}(2015){Paxton}, {Marchant}, {Schwab}, {Bauer},
  {Bildsten}, {Cantiello}, {Dessart}, {Farmer}, {Hu}, {Langer}, {Townsend},
  {Townsley}, \& {Timmes}}]{paxton2015}
{Paxton}, B., {Marchant}, P., {Schwab}, J., {et~al.} 2015, \apjs, 220, 15

\bibitem[{{Paxton} {et~al.}(2018){Paxton}, {Schwab}, {Bauer}, {Bildsten},
  {Blinnikov}, {Duffell}, {Farmer}, {Goldberg}, {Marchant}, {Sorokina},
  {Thoul}, {Townsend}, \& {Timmes}}]{paxton2018}
{Paxton}, B., {Schwab}, J., {Bauer}, E.~B., {et~al.} 2018, \apjs, 234, 34

\bibitem[{{Paxton} {et~al.}(2019){Paxton}, {Smolec}, {Schwab}, {Gautschy},
  {Bildsten}, {Cantiello}, {Dotter}, {Farmer}, {Goldberg}, {Jermyn}, {Kanbur},
  {Marchant}, {Thoul}, {Townsend}, {Wolf}, {Zhang}, \& {Timmes}}]{paxton2019}
{Paxton}, B., {Smolec}, R., {Schwab}, J., {et~al.} 2019, \apjs, 243, 10

\bibitem[{{Petersen}(1973)}]{Petersen1973}
{Petersen}, J.~O. 1973, \aap, 27, 89

\bibitem[{{Pols} {et~al.}(1995){Pols}, {Tout}, {Eggleton}, \& {Han}}]{Pols1995}
{Pols}, O.~R., {Tout}, C.~A., {Eggleton}, P.~P., \& {Han}, Z. 1995, \mnras,
  274, 964

\bibitem[{{Potekhin} \& {Chabrier}(2010)}]{Potekhin2010}
{Potekhin}, A.~Y. \& {Chabrier}, G. 2010, Contributions to Plasma Physics, 50,
  82

\bibitem[{{Przybilla} {et~al.}(2013){Przybilla}, {Nieva}, {Irrgang}, \&
  {Butler}}]{Przybilla2013}
{Przybilla}, N., {Nieva}, M.~F., {Irrgang}, A., \& {Butler}, K. 2013, in EAS
  Publications Series, Vol.~63, EAS Publications Series, ed. G.~{Alecian},
  Y.~{Lebreton}, O.~{Richard}, \& G.~{Vauclair}, 13--23

\bibitem[{{Ricker} {et~al.}(2015){Ricker}, {Winn}, {Vanderspek}, {Latham},
  {Bakos}, {Bean}, {Berta-Thompson}, {Brown}, {Buchhave}, {Butler}, {Butler},
  {Chaplin}, {Charbonneau}, {Christensen-Dalsgaard}, {Clampin}, {Deming},
  {Doty}, {De Lee}, {Dressing}, {Dunham}, {Endl}, {Fressin}, {Ge}, {Henning},
  {Holman}, {Howard}, {Ida}, {Jenkins}, {Jernigan}, {Johnson}, {Kaltenegger},
  {Kawai}, {Kjeldsen}, {Laughlin}, {Levine}, {Lin}, {Lissauer}, {MacQueen},
  {Marcy}, {McCullough}, {Morton}, {Narita}, {Paegert}, {Palle}, {Pepe},
  {Pepper}, {Quirrenbach}, {Rinehart}, {Sasselov}, {Sato}, {Seager},
  {Sozzetti}, {Stassun}, {Sullivan}, {Szentgyorgyi}, {Torres}, {Udry}, \&
  {Villasenor}}]{Ricker2015}
{Ricker}, G.~R., {Winn}, J.~N., {Vanderspek}, R., {et~al.} 2015, Journal of
  Astronomical Telescopes, Instruments, and Systems, 1, 014003

\bibitem[{{Rogers} \& {Nayfonov}(2002)}]{Rogers2002}
{Rogers}, F.~J. \& {Nayfonov}, A. 2002, \apj, 576, 1064

\bibitem[{{Saumon} {et~al.}(1995){Saumon}, {Chabrier}, \& {van
  Horn}}]{Saumon1995}
{Saumon}, D., {Chabrier}, G., \& {van Horn}, H.~M. 1995, \apjs, 99, 713

\bibitem[{{Seaton}(2005)}]{seaton2005}
{Seaton}, M.~J. 2005, \mnras, 362, L1

\bibitem[{{Stahler} {et~al.}(1980){Stahler}, {Shu}, \& {Taam}}]{Stahler1980}
{Stahler}, S.~W., {Shu}, F.~H., \& {Taam}, R.~E. 1980, \apj, 241, 637

\bibitem[{{Steindl} {et~al.}(2021{\natexlab{a}}){Steindl}, {Zwintz}, {Barnes},
  {M{\"u}llner}, \& {Vorobyov}}]{Steindl2021b}
{Steindl}, T., {Zwintz}, K., {Barnes}, T.~G., {M{\"u}llner}, M., \& {Vorobyov},
  E.~I. 2021{\natexlab{a}}, \aap, 654, A36

\bibitem[{{Steindl} {et~al.}(2021{\natexlab{b}}){Steindl}, {Zwintz}, \&
  {Bowman}}]{Steindl2020}
{Steindl}, T., {Zwintz}, K., \& {Bowman}, D.~M. 2021{\natexlab{b}}, \aap, 645,
  A119

\bibitem[{{Timmes} \& {Swesty}(2000)}]{Timmes2000}
{Timmes}, F.~X. \& {Swesty}, F.~D. 2000, \apjs, 126, 501

\bibitem[{Townsend(2020)}]{townsend2020}
Townsend, R. 2020, MESA SDK for Linux

\bibitem[{{Townsend} {et~al.}(2018){Townsend}, {Goldstein}, \&
  {Zweibel}}]{Townsend2018}
{Townsend}, R.~H.~D., {Goldstein}, J., \& {Zweibel}, E.~G. 2018, \mnras, 475,
  879

\bibitem[{{Townsend} \& {Teitler}(2013)}]{Townsend2013}
{Townsend}, R.~H.~D. \& {Teitler}, S.~A. 2013, \mnras, 435, 3406

\bibitem[{{Trotta}(2008)}]{Trotta2008}
{Trotta}, R. 2008, Contemporary Physics, 49, 71

\bibitem[{Van Der~Walt {et~al.}(2011)Van Der~Walt, Colbert, \&
  Varoquaux}]{van2011numpy}
Van Der~Walt, S., Colbert, S.~C., \& Varoquaux, G. 2011, Computing in Science
  \& Engineering, 13, 22

\bibitem[{{Virtanen} {et~al.}(2020){Virtanen}, {Gommers}, {Oliphant},
  {Haberland}, {Reddy}, {Cournapeau}, {Burovski}, {Peterson}, {Weckesser},
  {Bright}, {van der Walt}, {Brett}, {Wilson}, {Jarrod Millman}, {Mayorov},
  {Nelson}, {Jones}, {Kern}, {Larson}, {Carey}, {Polat}, {Feng}, {Moore}, {Vand
  erPlas}, {Laxalde}, {Perktold}, {Cimrman}, {Henriksen}, {Quintero}, {Harris},
  {Archibald}, {Ribeiro}, {Pedregosa}, {van Mulbregt}, \&
  {Contributors}}]{Virtanen_2020}
{Virtanen}, P., {Gommers}, R., {Oliphant}, T.~E., {et~al.} 2020, Nature
  Methods, 17, 261

\bibitem[{{Vorobyov} {et~al.}(2017){Vorobyov}, {Elbakyan}, {Hosokawa},
  {Sakurai}, {Guedel}, \& {Yorke}}]{Vorobyov2017}
{Vorobyov}, E.~I., {Elbakyan}, V., {Hosokawa}, T., {et~al.} 2017, \aap, 605,
  A77

\bibitem[{{Zwintz} {et~al.}(2014){Zwintz}, {Fossati}, {Ryabchikova},
  {Guenther}, {Aerts}, {Barnes}, {Theme{\ss}l}, {Lorenz}, {Cameron},
  {Kuschnig}, {Pollack-Drs}, {Moravveji}, {Baglin}, {Matthews}, {Moffat},
  {Poretti}, {Rainer}, {Rucinski}, {Sasselov}, \& {Weiss}}]{Zwintz2014}
{Zwintz}, K., {Fossati}, L., {Ryabchikova}, T., {et~al.} 2014, Science, 345,
  550

\end{thebibliography}

\begin{appendix}

\section{MESA microphysics}
\label{app:mesaphysics}
The MESA EOS is a blend of the OPAL \citet{Rogers2002}, SCVH
\citet{Saumon1995}, PTEH \citet{Pols1995}, HELM
\citet{Timmes2000}, and PC \citet{Potekhin2010} EOSes.
Radiative opacities are primarily from OPAL \citep{Iglesias1993,
Iglesias1996}, with low-temperature data from \citet{Ferguson2005}
and the high-temperature, Compton-scattering-dominated regime by
\citet{Buchler1976}.  Electron conduction opacities are from
\citet{Cassisi2007}.

Nuclear reaction rates are a combination of rates from
NACRE \citep{Angulo1999}, JINA REACLIB \citep{Cyburt2010}, and
additional tabulated weak reaction rates. (\citet{Fuller1985, Oda1994,
Langanke2000}. Screening
is included via the prescription of \citet{Chugunov2007}.  Thermal
neutrino loss rates are from \citet{Itoh1996}.

\FloatBarrier
\section{Estimating the stellar parameter of theoretical echelle diagrams}

  \begin{table}
\begin{footnotesize}
    \caption[]{Extracted stellar parameters of model \#10001 for the different approaches with a Rayleigh limit of a 108-day light curve.}
    \label{tab:paramet_model10001_107d}
    \tabcolsep=0.04cm
    \begin{tabular*}{\linewidth}{lcccccc}
        \hline
        \noalign{\smallskip}
                & \#10001   & $\chi^2$  & $\chi^2$      & $\chi^2$      & $\chi^2_{\rm spec}$    & $\chi^2_{\rm spec}$  \\
                &           &           & $3\sigma$     &$ 1\sigma$     &        $\leq 1$        &   $\leq 28.87$       \\
        \noalign{\smallskip}
        \hline
        \noalign{\smallskip}
        \multicolumn{7}{l}{spectroscopic parameters assumed correct} \\
        $M_\star$ ($M_\odot$)    &   1.849   &   2.11(18) & \bf{1.85(5)} & 1.81(3)  & 1.81(2)  & \bf{1.76(9)} \\
        $T_{\rm eff}$ (K)    &   7431   &   8800(800) & 7680(200) & \bf{7460(80)}  & \bf{7430(30)}  & \bf{7480(130)} \\
        $\log\,g$    &   4.208   &   4.225(12) & \bf{4.208(4)} & \bf{4.205(3)}  & 4.205(2)  & \bf{4.200(8)} \\
        $R$ ($R_\odot$)&   1.772   &  1.85(5) & \bf{1.77(2)} & 1.76(1)  & 1.759(6)  & \bf{1.75(3)} \\
        $L$ ($L_\odot$)&   8.6  &  19(7) & \bf{9.9(1.2)} & \bf{8.6(4)}  & \bf{8.5(2)}  & \bf{8.6(7)} \\
        age (Myr)&   8.12  &  5.9(1.6) & \bf{7.8(1.4)} & \bf{8.1(8)}  & \bf{8.4(7)}  & \bf{8.8(1.4)} \\
        Z&   0.018 &  \bf{0.018(3)} & \bf{0.018(2)} & \bf{0.018(2)}  & 0.020(1)  & \bf{0.018(4)} \\
        \noalign{\smallskip}
        \hline
        \multicolumn{7}{l}{spectroscopic parameters assumed $T_{\rm eff} = 7488$\,K and $\log\,g=4.275$} \\
        $M_\star$ ($M_\odot$)    &   1.849   &   2.10(18) & \bf{1.86(6)} & \bf{1.82(3)}  & 1.79(4)  & \bf{1.77(10)} \\
        $T_{\rm eff}$ (K)    &   7431   &   8700(700) & 7720(230) & \bf{7500(110)}  & 7480(30)  & \bf{7510(140)} \\
        $\log\,g$    &   4.208   &   4.225(12) & \bf{4.208(4)} & \bf{4.205(3)}  & 4.203(3)  & \bf{4.200(9)} \\
        $R$ ($R_\odot$)&   1.772   &  1.85(5) & \bf{1.78(2)} & \bf{1.762(11)}  & 1.75(2)  & \bf{1.75(3)} \\
        $L$ ($L_\odot$)&   8.6  &  20(8) & 10.1(1.3) & \bf{8.9(6)}  & \bf{8.66(13)}  & \bf{8.8(8)} \\
        age (Myr)&   8.12  &  6.1(1.6) & \bf{7.8(1.4)} & \bf{8.2(1.1)}  & \bf{8.4(4)}  & \bf{8.7(1.4)} \\
        Z&   0.018 &  \bf{0.018(3)} & \bf{0.018(2)} & \bf{0.018(2)}  & \bf{0.018(1)}  & \bf{0.018(4)} \\
        \noalign{\smallskip}
        \hline
        \multicolumn{7}{l}{spectroscopic parameters assumed $T_{\rm eff} = 7553$\,K and $\log\,g=4.127$} \\
        $M_\star$ ($M_\odot$)    &   1.849   &   2.13(18) & \bf{1.88(8)} & \bf{1.83(5)}  & -  & \bf{1.78(10)} \\
        $T_{\rm eff}$ (K)    &   7431   &   8900(800) & 7770(250) & 7560(120)  & -  & \bf{7560(140)} \\
        $\log\,g$    &   4.208   &   4.227(12) & \bf{4.209(6)} & \bf{4.206(4)}  & -  & \bf{4.20(9)} \\
        $R$ ($R_\odot$)&   1.772   &  1.86(5) & \bf{1.78(2)} & 1.767(15)  & -  & \bf{1.75(3)} \\
        $L$ ($L_\odot$)&   8.6  &  21(8) & 10.5(1.6) & \bf{9.2(7)}  & - & \bf{9.0(9)} \\
        age (Myr)&   8.12  &  \bf{5.6(1.6)} & \bf{7.5(1.3)} & \bf{8.1(1.1)}  &-  & \bf{8.5(1.4)} \\
        Z&   0.018 &  \bf{0.018(3)} & \bf{0.018(2)} & \bf{0.018(1)}  & -  & \bf{0.018(4)} \\
        \noalign{\smallskip}
        \hline
        \multicolumn{7}{l}{spectroscopic parameters assumed $T_{\rm eff} = 7261$\,K and $\log\,g=4.145$} \\
        $M_\star$ ($M_\odot$)    &   1.849   &   2.11(18) & \bf{1.83(5)} & 1.80(4)  & -  & 1.74(9) \\
        $T_{\rm eff}$ (K)    &   7431   &   8800(800) & \bf{7580(160)} & \bf{7400(40)}  & -  & \bf{7370(110)} \\
        $\log\,g$    &   4.208   &   4.225(12) & \bf{4.206(4)} & 4.204(3)  & -  & 4.198(9) \\
        $R$ ($R_\odot$)&   1.772   &  1.85(6) & \bf{1.77(2)} & 1.76(1)  & -  & 1.74(3) \\
        $L$ ($L_\odot$)&   8.6  &  19(7) & \bf{9.3(9)} & 8.3(2)  & -  & \bf{8.0(6)} \\
        age (Myr)&   8.12  &  6.0(1.6) & \bf{8.1(1.4)} & \bf{8.6(8)}  & -  & \bf{9.3(1.5)} \\
        Z&   0.018 &  \bf{0.018(3)} & \bf{0.018(2)} & \bf{0.020(1)}  & -  & \bf{0.018(4)} \\
        \noalign{\smallskip}
        \hline
        \multicolumn{7}{l}{spectroscopic parameters assumed $T_{\rm eff} = 7368$\,K and $\log\,g=4.17$} \\
        $M_\star$ ($M_\odot$)    &   1.849   &   2.11(18) & \bf{1.84(5)} & 1.80(4)  & 1.78(4)  & \bf{1.76(9)} \\
        $T_{\rm eff}$ (K)    &   7431   &   8800(800) & 7640(190) & \bf{7430(60)}  & 7380(30)  & \bf{7440(120)} \\
        $\log\,g$    &   4.208   &   4.225(12) & \bf{4.207(4)} & \bf{4.204(4)}  & 4.202(4)  & 4.199(8) \\
        $R$ ($R_\odot$)&   1.772   &  1.85(6) & \bf{1.77(2)} & 1.754(13)  & 1.750(13)  & \bf{1.74(4)} \\
        $L$ ($L_\odot$)&   8.6  &  20(8) & \bf{9.7(1.1)} & \bf{8.4(3)}  & 8.2(2)  & \bf{8.4(7)} \\
        age (Myr)&   8.12  &  \bf{6.0(1.6)} & \bf{8.1(1.4)} & \bf{8.5(7)}  & 8.6(8)  & \bf{9.0(1.4)} \\
        Z&   0.018 &  \bf{0.018(3)} & \bf{0.018(2)} & \bf{0.02(2)}  & \bf{0.02(2)}  & \bf{0.018(4)} \\
        \noalign{\smallskip}
        \hline
        \multicolumn{7}{l}{correct percentage} \\
        \%&   - & 20 & 86 & 74 & 20 & 89 \\
        \noalign{\smallskip}
        \hline

    \end{tabular*}    
    \tablefoot{The expectation value of the extracted parameters adopted as the mean of all chosen models. Only for the metallicity, Z, did we adopt the median as the expectation value since the grid is evenly spaced. Values in parentheses give the uncertainties on the extracted parameters adopted from the standard deviation of all models. Parameters that agree with the model are given in bold. Values are not given if there are fewer than five models.
    }
\end{footnotesize}
 \end{table}

\begin{table}
    \caption[]{Percentage rate of extracted stellar parameters that agree with the model's values within $3\sigma$.}
    \label{tab:all_approaches_overview_3sigma}
    \begin{tabular}{lp{1cm}p{1cm}p{1cm}p{1cm}c}
        \hline
        \noalign{\smallskip}
                   & $\chi^2$  & $\chi^2$      & $\chi^2$      & $\chi^2_{\rm spec}$    & $\chi^2_{\rm spec}$  \\
                   &           & $3\sigma$     &$ 1\sigma$     &        $\leq 1$        &   $\leq 28.87$       \\
                   &     \%      & \%     & \%      &        \%         &   \%        \\
        \noalign{\smallskip}
        \hline
        \noalign{\smallskip}
        \multicolumn{6}{l}{Rayleigh limit corresponding to 27 day light curve} \\
        $M_\star$    & 100 & 100 & 78 & 94 & 99 \\
        $T_{\rm eff}$        & 100 & 100 & 70 & 40 & 97 \\
        $\log\,g$               & 100 & 100 & 73 & 93 & 99 \\
        $R$          & 100 & 100 & 79 & 93 & 100 \\
        $L$         & 100 & 100 & 73 & 63 & 97 \\
        age               & 97  & 99  & 76 & 94 & 99 \\
        Z                       & 100 & 100 & 77 & 97 & 100 \\
        \noalign{\smallskip}
        \hline\noalign{\smallskip}
        total                   & 100 & 100 & 75 & 82 & 99 \\
        \noalign{\smallskip}
        \hline\noalign{\smallskip}
        \multicolumn{6}{l}{Rayleigh limit corresponding to 108 day light curve} \\
        $M_\star$    & 100 & 100 & 71 & 83 & 99 \\
        $T_{\rm eff}$       & 100 & 100 & 71 & 40 & 97 \\
        $\log\,g$               & 100 & 100 & 69 & 79 & 99 \\
        $R$          & 100 & 100 & 73 & 81 & 100 \\
        $L$         & 100 & 100 & 69 & 51 & 94 \\
        age               & 100 & 100 & 76 & 80 & 99 \\
        Z                       & 100 & 100 & 77 & 84 & 100 \\
        \noalign{\smallskip}
        \hline\noalign{\smallskip}
        total                   & 100 & 100 & 72 & 71 & 98 \\
        \noalign{\smallskip}
        \hline

    \end{tabular}      
 \end{table}

\begin{table}
    \caption[]{Median $1\sigma$ uncertainty on the extracted parameters.}
    \label{tab:all_approaches_overview_uncertainty}
    \begin{tabular}{lp{1cm}p{1cm}p{1cm}p{1cm}c}
        \hline
        \noalign{\smallskip}
                   & $\chi^2$  & $\chi^2$      & $\chi^2$      & $\chi^2_{\rm spec}$    & $\chi^2_{\rm spec}$  \\
                   &           & $3\sigma$     &$ 1\sigma$     &        $\leq 1$        &   $\leq 28.87$       \\
                   &     \%      & \%     & \%      &        \%         &   \%        \\
        \noalign{\smallskip}
        \hline
        \noalign{\smallskip}
        \multicolumn{6}{l}{Rayleigh limit corresponding to 27 day light curve} \\
        $M_\star$ ($M_\odot$)   & 0.19 & 0.11 & 0.08 & 0.07 & 0.13 \\
        $T_{\rm eff}$ (K)       & 700 & 300 & 110 & 28 & 150 \\
        $\log\,g$               & 0.014 & 0.008 & 0.006 & 0.005 & 0.015 \\
        $R$ ($R_\odot$)         & 0.06 & 0.04 & 0.03 & 0.0325 & 0.05 \\
        $L$ ($L_\odot$)         & 8 & 3.4 & 1.4 & 0.6 & 2.0 \\
        age (Myr)               & 1.7  & 1.3  & 0.85 & 0.75 & 1.7 \\
        Z                       & 0.005 & 0.004 & 0.004 & 0.004 & 0.004 \\
        \noalign{\smallskip}
        \hline\noalign{\smallskip}
        \multicolumn{6}{l}{Rayleigh limit corresponding to 108 day light curve} \\
        $M_\star$ ($M_\odot$)   & 0.13 & 0.08 & 0.06 & 0.05 & 0.11 \\
        $T_{\rm eff}$ (K)       & 600 & 300 & 110 & 25 & 150 \\
        $\log\,g$               & 0.010 & 0.006 & 0.005 & 0.004 & 0.009 \\
        $R$ ($R_\odot$)         & 0.04 & 0.03 & 0.02 & 0.017 & 0.04 \\
        $L$ ($L_\odot$)         & 6.5 & 3.1 & 1.2 & 0.4 & 1.8 \\
        age (Myr)               & 1.2 & 0.9 & 0.7 & 0.5 & 1.6 \\
        Z                       & 0.004 & 0.003 & 0.003 & 0.003 & 0.005 \\
        \noalign{\smallskip}
        \noalign{\smallskip}
        \hline

    \end{tabular}  
    \tablefoot{We note that for approaches 3 and 4 it is possible that not a single model fulfils both the stringent spectroscopic and the asteroseismic demands. The values given here are the medians of the models for which at least two theoretical pulsation spectra from different evolutionary calculations fit both constraints.
    }
 \end{table}

\FloatBarrier
\section{Petersen diagrams}

 \begin{figure}
   \centering
   \includegraphics[width=\linewidth]{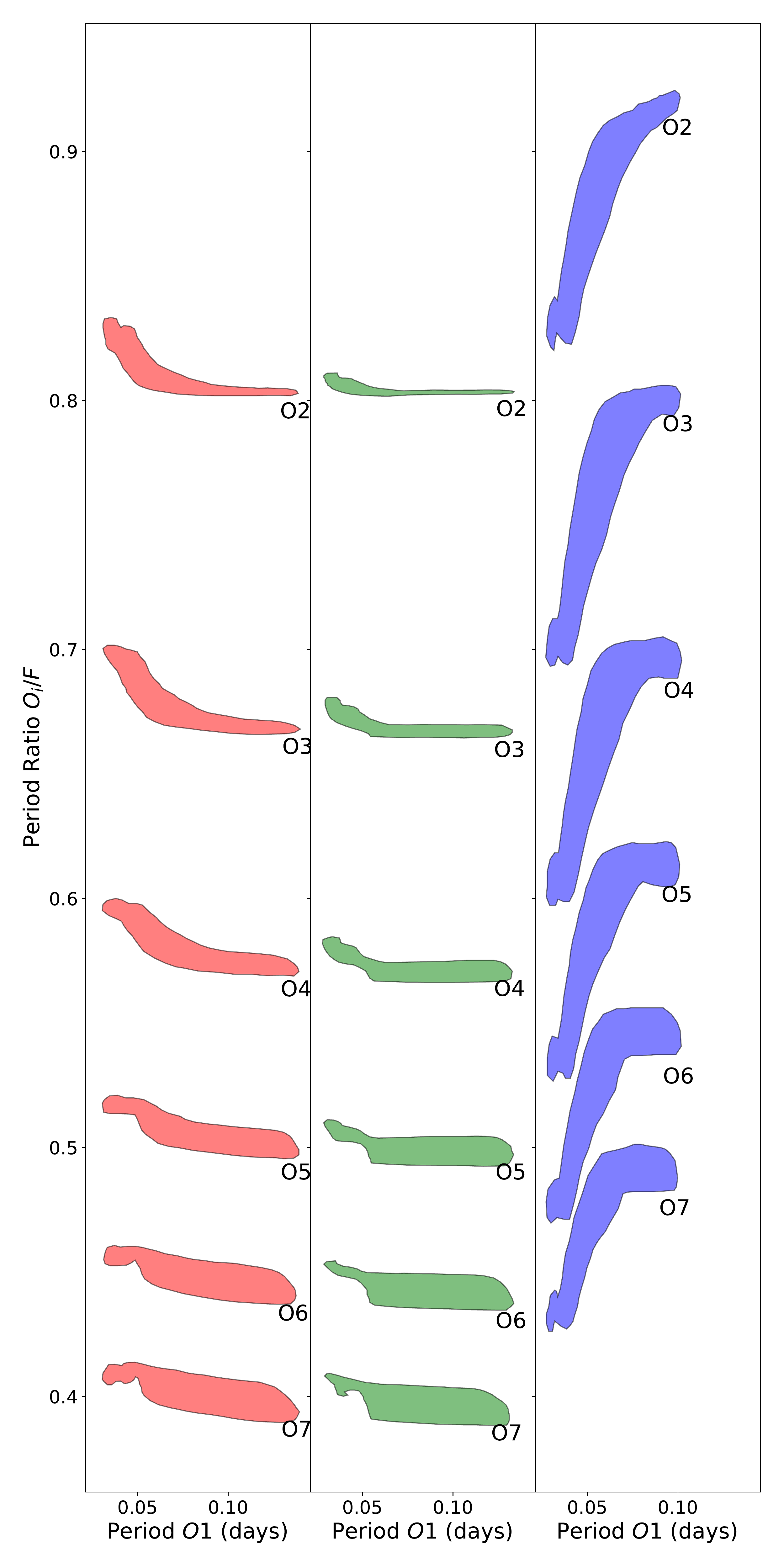}
      \caption{Peterson diagram for the first overtone.
              }
         \label{fig:PetersenO1}
\end{figure}
 \begin{figure}
   \centering
   \includegraphics[width=\linewidth]{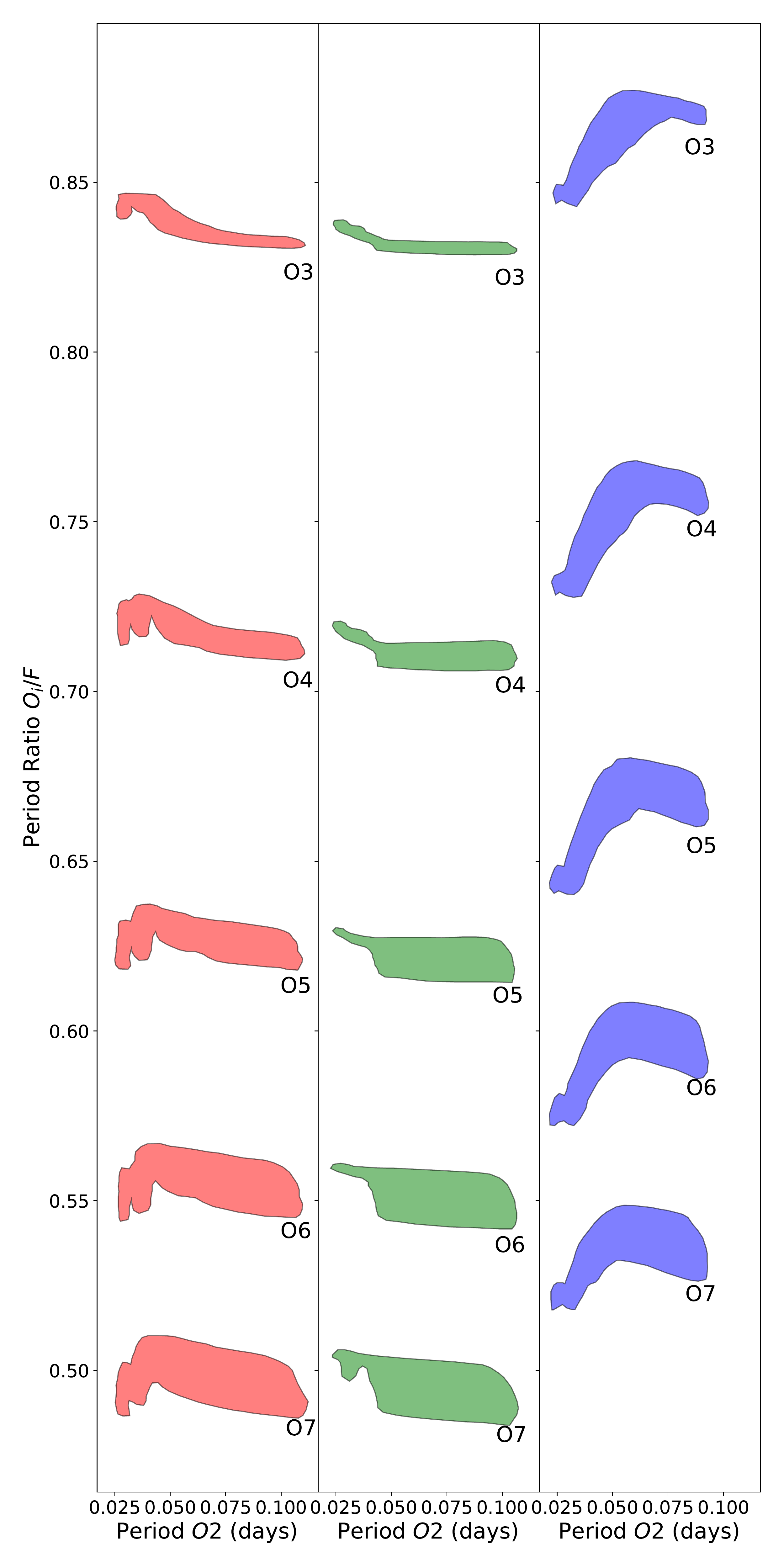}
      \caption{Peterson diagram for the second overtone.
              }
         \label{fig:PetersenO2}
\end{figure}
 \begin{figure}
   \centering
   \includegraphics[width=\linewidth]{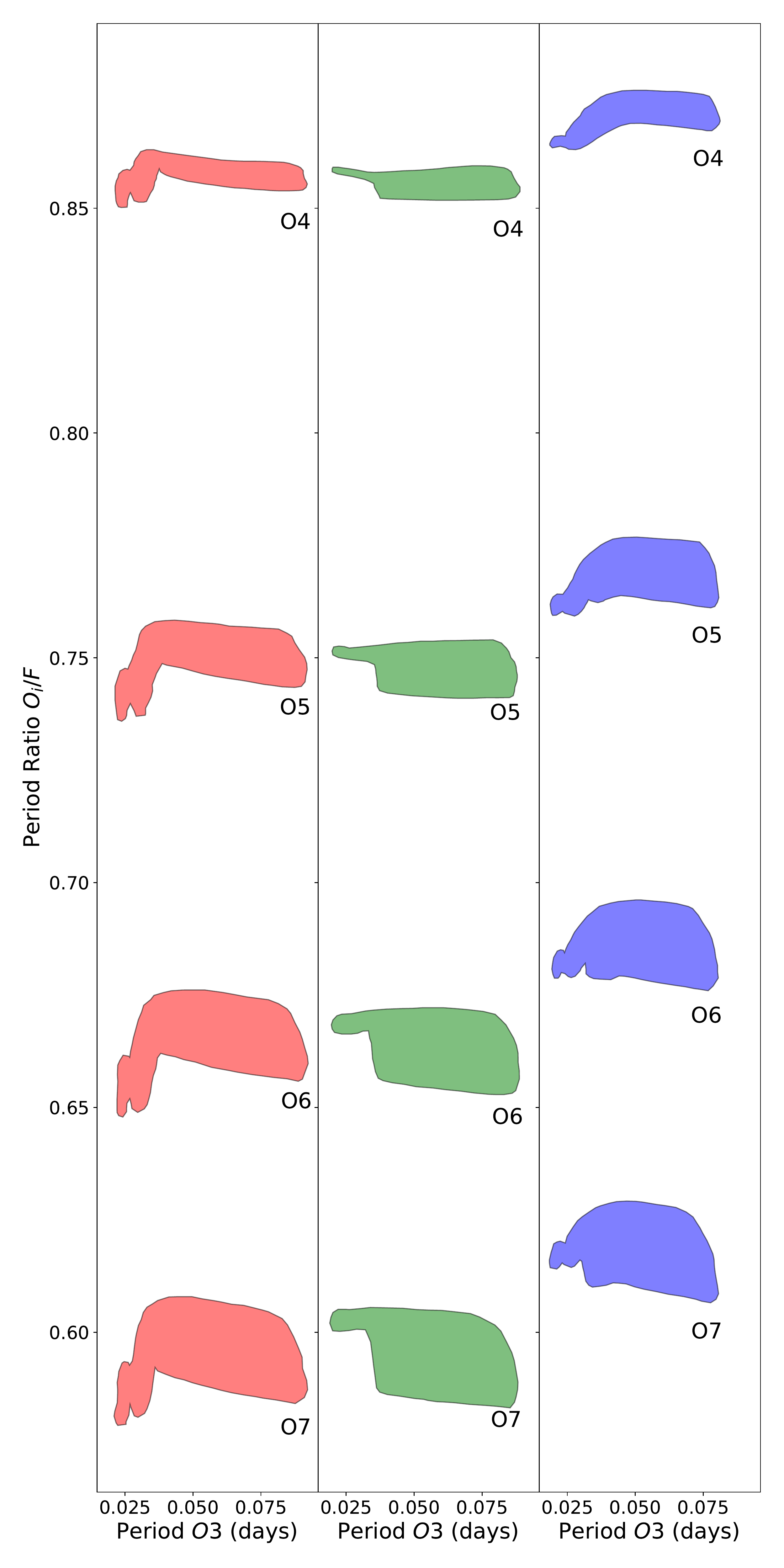}
      \caption{Peterson diagram for the third overtone.
              }
         \label{fig:PetersenO3}
\end{figure}

 \begin{figure}
   \centering
   \includegraphics[width=\linewidth]{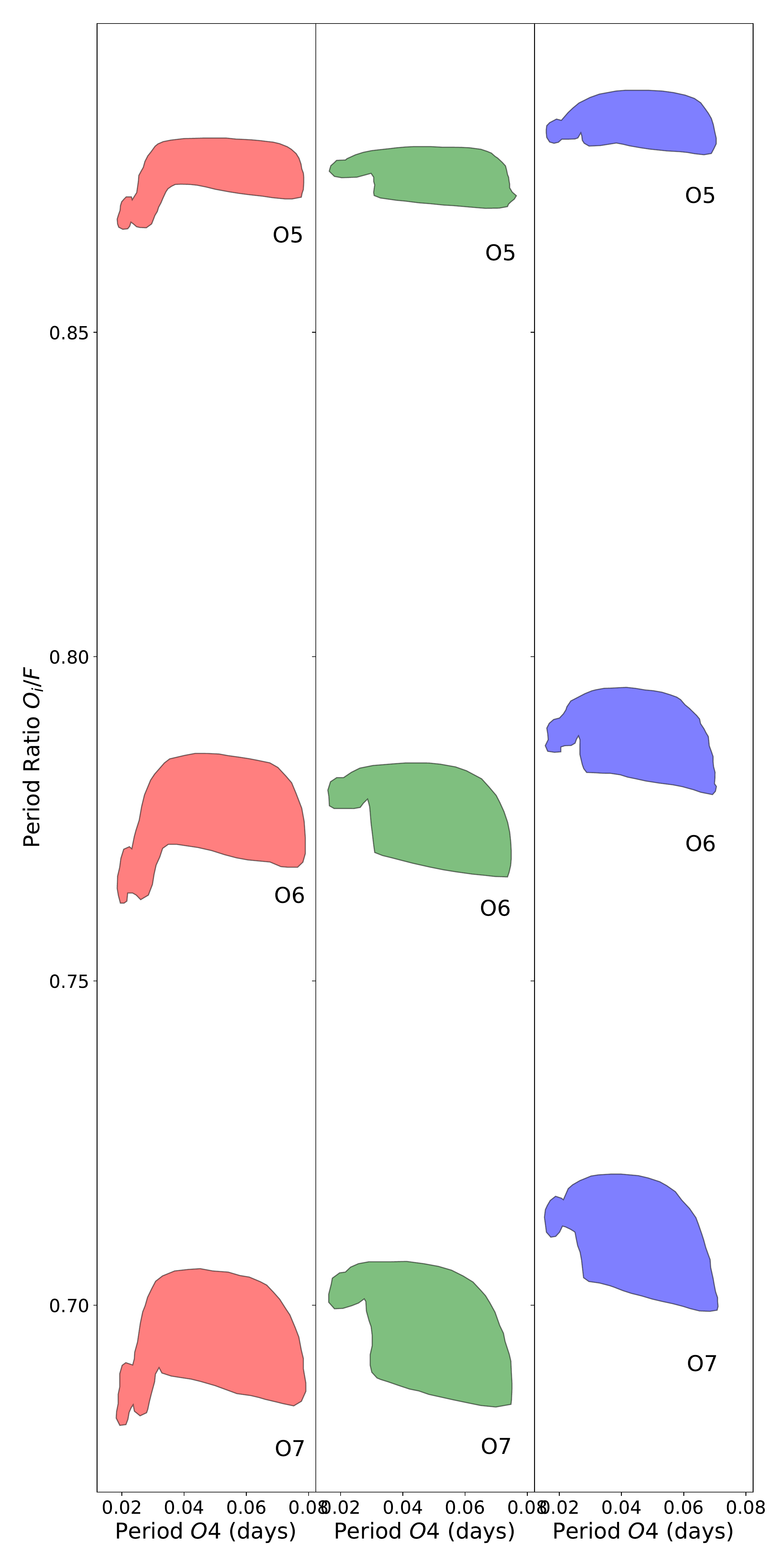}
      \caption{Peterson diagram for the fourth overtone.
              }
         \label{fig:PetersenO4}
\end{figure}

\FloatBarrier
\section{Re-discussion of HD 139614}

 \begin{figure*}
   \centering
   \includegraphics[width=\linewidth]{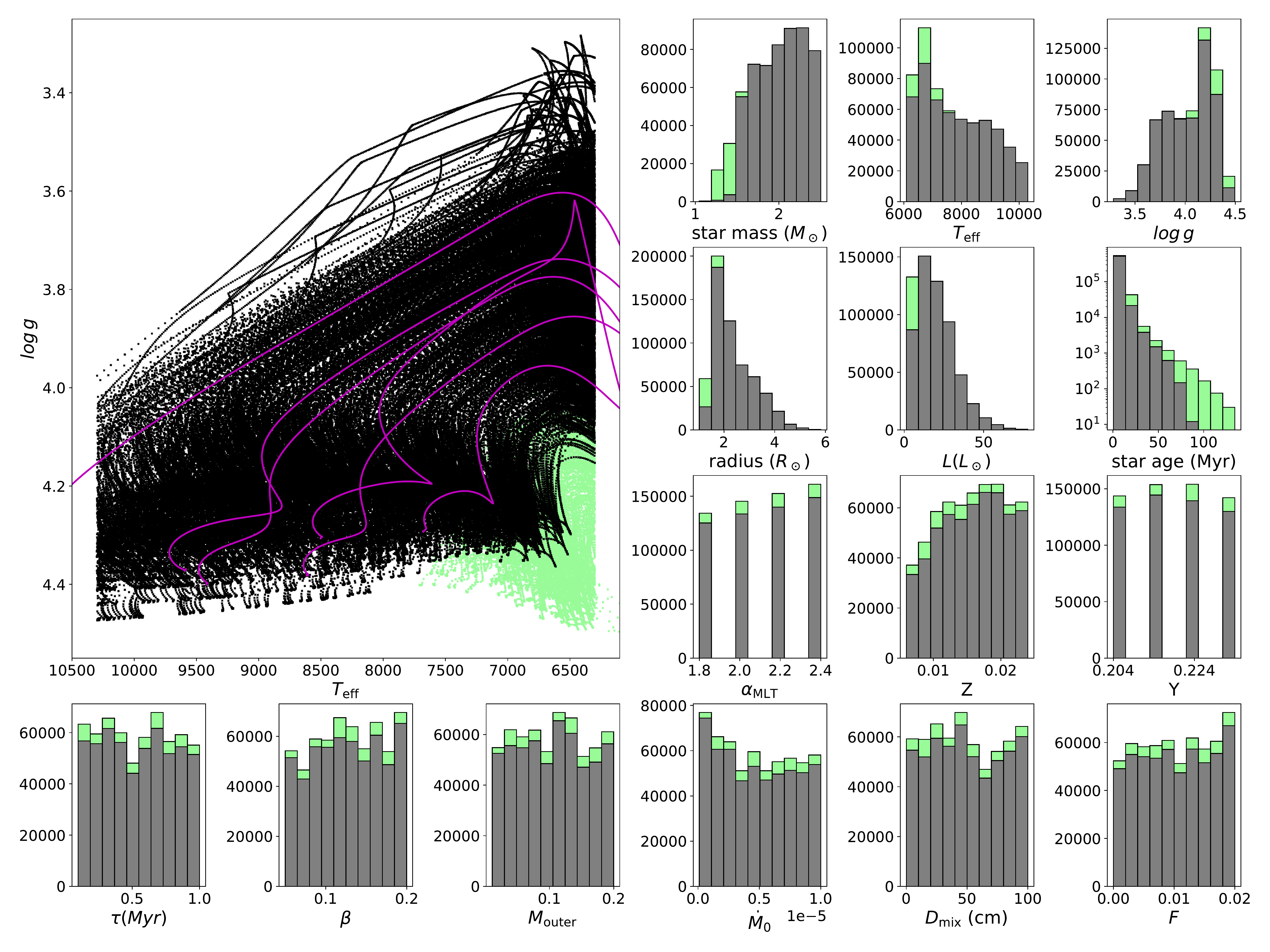}
      \caption{Similar to Figure \ref{fig:parameter_overview}, but including the extension for the modelling of HD 139614. The additional models calculated are added to the Kiel diagram and stacked upon the histograms in pale green. 
              }
         \label{fig:parameter_overview_extended}
\end{figure*}

\begin{table}
\begin{footnotesize}
    \caption[]{Extracted stellar parameters of HD 139614 when applying the six-dimensional model grid.}
    \label{tab:paramet_rediscussion_6dim}
    \tabcolsep=0.04cm
    \begin{tabular*}{\linewidth}{lcccccc}
        \hline
        \noalign{\smallskip}
                & \citet{murphy2021}    & $\chi^2$      & $\chi^2$      & $\chi^2_{\rm spec}$  \\
                &  modelling               & $3\sigma$     &$ 1\sigma$     &   $\leq 22.36$       \\
        \noalign{\smallskip}
        \hline
        \noalign{\smallskip}
        \multicolumn{7}{l}{same mode identification as in \citet{murphy2021}} \\
        $M_\star$ ($M_\odot$)    &   1.520(18)   &  1.71(11) & 1.67(9)  & 1.66(10)   \\
        $T_{\rm eff}$ (K)    &   7615(64)   &   7840(260) & 7670(120) & 7660(130)  \\
        $\log\,g$    &   -   &   4.289(9) & 4.286(8) & 4.283(9)\\
        $R$ ($R_\odot$)&   -   &  1.55(3) & 1.54(3) & 1.54(4)  \\
        $L$ ($L_\odot$)&   6.7(3)  &  8.2(1.3) & 7.4(6)  & 7.4(7) \\
        age (Myr)&   10.75(77)  &  12(3) & 13(3) & 12(3)  \\
        Z&   0.010(1) &  0.018(5) & 0.016(4) & 0.014(5)\\
        \noalign{\smallskip}
        \hline

    \end{tabular*}    
    \tablefoot{The expectation value of the extracted parameter adopted as the mean of all chosen models. Only for the metallicity, Z, did we adopt the median as the expectation value since the grid is evenly spaced. Values in parentheses give the uncertainty on the extracted parameter adopted from the standard deviation of all models. We note that this model grid consists of about a factor of three fewer model calculations than than the ten-dimensional model grid. 
    }
\end{footnotesize}
 \end{table}
 
\end{appendix}

\end{document}